\documentclass{aa}  
\usepackage{amsmath}
\usepackage{graphicx}
\usepackage{graphics}
\usepackage{subfigure}
\usepackage{txfonts}
\usepackage{natbib}
\usepackage{xcolor}
\usepackage{soul}
\usepackage{tablefootnote}

\bibpunct{(}{)}{;}{a}{}{,}

\def\teff{\ifmmode T_{\rm eff} \else $T_{\mathrm{eff}}$\fi}

\def\ltsima{$\buildrel<\over\sim$}
\def\lsim{\lower.5ex\hbox{\ltsima}}

\newcommand{\ha}{\ifmmode {\rm H}\alpha \else H$\alpha$\fi}
\newcommand{\hb}{\ifmmode {\rm H}\beta \else H$\beta$\fi}
\newcommand{\lya}{\ifmmode {\rm Ly}\alpha \else Ly$\alpha$\fi}

\newcommand{\heii}{He~{\sc ii}}

\newcommand{\ebv}{\ifmmode E_{\rm B-V} \else $E_{\rm B-V}$\fi}
\newcommand{\av}{\ifmmode A_{\rm V} \else $A_{\rm V}$\fi}
\def\msun{\ifmmode M_{\odot} \else M$_{\odot}$\fi}
\def\msunyr{\ifmmode M_{\odot} {\rm yr}^{-1} \else M$_{\odot}$ yr$^{-1}$\fi}
\def\zsun{\ifmmode Z_{\odot} \else Z$_{\odot}$\fi}
\def\lsun{\ifmmode L_{\odot} \else L$_{\odot}$\fi}

\def\mup{\ifmmode M_{\rm up} \else M$_{\rm up}$\fi}
\def\mlow{\ifmmode M_{\rm low} \else M$_{\rm low}$\fi}

\def\aap{A\&A}

\def\apjl{ApJL}
\def\apjs{ApJS}

\newcommand{\oh}{\ifmmode 12 + \log({\rm O/H}) \else$12 + \log({\rm O/H})$\fi}
\newcommand{\nii}{[N~{\sc ii}]}
\newcommand{\oi}{[O~{\sc i}]}
\newcommand{\oii}{[O~{\sc ii}]}
\newcommand{\sii}{[S~{\sc ii}]}
\newcommand{\siii}{[S~{\sc iii}]}
\newcommand{\oiii}{[O~{\sc iii}]}
\def\Sii{[S~{\sc ii}] $\lambda\lambda$6716,6731}
\def\Siii{[S~{\sc iii}] $\lambda$9068} %\lambda 9532 LR: I use only [SIII]9069 because the other in not in DR14
\def\Oi{[O~{\sc i}] $\lambda$6300}
\def\Oii{[O~{\sc ii}] $\lambda\lambda$3726, 3728}
\def\Oiii{[O~{\sc iii}] $\lambda\lambda$4959,5007}
\def\oiiil{[O~{\sc iii}]$\lambda 5007$}
\def\flyf{\ifmmode f_{\rm Lyf} \else $f_{\rm Lyf}$\fi}
\def\pz{\ifmmode P(z) \else $P(z)$\fi}
\def\ki2{\ifmmode \chi^2 \else $\chi^2$\fi}
\def\zphot{\ifmmode z_{\rm phot} \else $z_{\rm phot}$\fi}

\newcommand{\xphot}{\ifmmode x_\gamma \else $v_\gamma$\fi}
\newcommand{\xobs}{\ifmmode x_{\rm obs} \else $x_{\rm obs}$\fi}
\newcommand{\xcmf}{\ifmmode x_{\rm CMF} \else $x_{\rm CMF}$\fi}
\newcommand{\vexp}{\ifmmode V_{\rm exp} \else $V_{\rm exp}$\fi}
\newcommand{\vmax}{\ifmmode V_{\rm max} \else $V_{\rm max}$\fi}
\newcommand{\nh}{\ifmmode N_{\rm HI} \else $N_{\rm HI}$\fi}
\newcommand{\dv}{\ifmmode \Delta v({\rm em-abs}) \else $\Delta v({\rm em}-{\rm abs})$\fi}

\def\fesc{\ifmmode f_{\rm esc} \else $f_{\rm esc}$\fi}

\def\frellya{\ifmmode f^{\rm rel}_{\rm{Ly}\alpha} \else $f^{\rm rel}_{\rm{Ly}\alpha}$\fi}

\newcommand{\mstar}{\ifmmode M_\star \else $M_\star$\fi}
\newcommand{\muv}{\ifmmode M_{1500} \else $M_{1500}$\fi}
\newcommand{\auv}{\ifmmode A_{\rm UV} \else $A_{\rm UV}$\fi}
\newcommand{\luv}{\ifmmode L_{\rm UV} \else $L_{\rm UV}$\fi}
\newcommand{\lir}{\ifmmode L_{\rm IR} \else $L_{\rm IR}$\fi}
\newcommand{\lbol}{\ifmmode L_{\rm bol} \else $L_{\rm bol}$\fi}
\newcommand{\liruv}{\ifmmode L_{\rm IR+UV} \else $L_{\rm IR+UV}$\fi}
\newcommand{\liroveruv}{\ifmmode L_{\rm IR}/L_{\rm UV} \else $L_{\rm IR}/L_{\rm UV}$\fi}
\newcommand{\nlyc}{\ifmmode N_{\rm Lyc} \else $N_{\rm Lyc} $\fi}
\newcommand{\rholyc}{\ifmmode \rho_{\rm Lyc} \else $\rho_{\rm Lyc} $\fi}
\newcommand{\chion}{\ifmmode \xi_{\rm ion} \else $\xi_{\rm ion}$\fi}
\newcommand{\chioncorr}{\ifmmode \xi_{\rm ion}^0 \else $\xi_{\rm ion}^0$\fi}

\newcommand{\uhigh}{\ifmmode U_{\rm high} \else $U_{\rm high}$\fi}
\newcommand{\ulow}{\ifmmode U_{\rm low} \else $U_{\rm low}$\fi}

\begin{document}

    \title{Reconciling escape fractions and observed line emission in Lyman-continuum-leaking galaxies
    \thanks{Based on data obtained with the European Southern Observatory Very Large Telescope, Paranal, Chile, under 0102.B-0942(A)}}
  \subtitle{}
  \author{L. Ramambason\inst{1,2}, D. Schaerer\inst{1,3}, 
G. Stasi\'nska$^{4}$, 
Y. I. Izotov$^{5}$,
N. G. Guseva$^{5}$,
J.M V\'ilchez$^{6}$,
R. Amor\'in$^{7,8}$,
C. Morisset$^9$}
%  \offprints{}
  \institute{Observatoire de Gen\`eve, Universit\'e de Gen\`eve, 51 Ch. des Maillettes, 1290 Versoix, Switzerland
         \and
AIM, CEA, Université Paris-Saclay, Université Paris Diderot, Université de Paris, F-91191, Gif-sur-Yvette, France.
         \and
CNRS, IRAP, 14 Avenue E. Belin, 31400 Toulouse, France
        \and
LUTH, Observatoire de Meudon, 92195 Meudon Cedex, France
        \and
Bogolyubov Institute for Theoretical Physics, National Academy of Sciences of Ukraine, 14-b Metrolohichna str., Kyiv,
03143, Ukraine
    \and
Instituto de Astrofísica de Andalucía, CSIC, Apartado de correos 3004, E-18080 Granada, Spain
    \and
Instituto de Investigación Multidisciplinar en Ciencia y Tecnología, Universidad de La Serena, Raúl Bitrán 1305, La Serena, Chile
    \and
Departamento de Astronomía, Universidad de La Serena, Av. Juan Cisternas 1200 Norte, La Serena, Chile
\and
Instituto de Astronomía, Universidad Nacional Autónoma de México, AP 106, 22800 Ensenada, B. C., Mexico
}

\date{Received 11 June 2020 / Accepted 17 September 2020}
%\date{Accepted for publication in A\&A}

\abstract{%CONTEXT
Finding and elucidating the properties of Lyman-continuum(LyC)-emitting galaxies is an important step in improving our understanding of cosmic reionization.}
{%AIMS
Although the $z \sim 0.3-0.4$ LyC emitters found recently show strong optical emission lines, no consistent quantitative photoionization model taking 
into account the escape of ionizing photons and inhomogenous interstellar medium (ISM) geometry of these galaxies has yet been constructed. 
Furthermore, it is unclear to what extent these emission lines can be used to distinguish LyC emitters.
}
{%METHODS
To address these questions we construct one- and two-zone photoionization models accounting for the observed LyC escape, which we compare to the observed
emission line measurements. The main diagnostics used include lines of \oiii, \oii, and \oi\  plus sulfur lines (\sii, \siii) and a nitrogen line (\nii), which probe regions of 
different ionization in the ISM.
}
{%RESULTS
We find that single (one-zone) density-bounded photoionization models cannot reproduce the emission lines of the LyC leakers, as pointed out by earlier studies, 
because they systematically underpredict the lines of species of low ionization potential, such as \oi\ and \sii. 
Introducing a two-zone model, with differing ionization parameter and a variable covering fraction and where one of the zones is density-bounded,
we show that the observed emission line ratios of the LyC emitters are well reproduced. 
Furthermore, our model yields LyC escape fractions, which are in fair agreement with the observations and
independent measurements.
The \Oi\ excess, which is observed in some LyC leakers, can be naturally explained in this model, for example\ by emission from low-ionization and low-filling-factor gas.
LyC emitters with a high escape fraction (\fesc$\protect\ga38$\%) are deficient both in \Oi\ and in \Sii.
We also confirm that a \Sii\ deficiency can be used to select LyC emitter candidates, as suggested earlier.
Finally, we find indications for a possible dichotomy in terms of escape mechanisms for LyC photons between galaxies with relatively low (\fesc$\protect\la10$\%)
and higher escape fractions.}
{%CONCLUSIONS
We conclude that two-zone photoionization models are sufficient and required to explain the observed emission line properties of $z \sim 0.3-0.4$ LyC emitters.
This is in agreement with UV absorption line studies, which also show the co-existence of regions with  high hydrogen column density (i.e.,\ no escape
of ionizing photons) and density-bounded or very low column density regions responsible for the observed escape of LyC radiation. These simple but consistent models
provide a first step towards the use of optical emission lines and their ratios as quantitative diagnostics of LyC escape from galaxies.}
% 5 {} token are mandatory

 \keywords{Galaxies: starburst -- Galaxies: high-redshift -- Cosmology: dark ages, reionization, first stars 
 -- Ultraviolet: galaxies}

\authorrunning{L. Ramambason et al.}
\titlerunning{Consistent emission line modeling of Lyman continuum emitters}
\maketitle

%%%%%%%%%%%%%%%%%%%%%%%%%%%%%%%%%%%%%%%%%%%%%%%%%%%%%%%%%%%%%%%%%%%%%%%%%%%%%%%%%s
\section{Introduction}
\label{s_intro}

Significant progress has been made in recent years in searches for and studies of star-forming galaxies, which emit copious amounts of Lyman continuum (LyC) radiation. This ionizing radiation can escape the interstellar medium (ISM) and ionize the surroundings, thus contributing to intergalactic ionizing radiation. If a sufficient population of such galaxies exists in the early Universe, their presence could explain cosmic reionization, which is known to have started shortly after the Big Bang and to have ended $\sim 1$ Gyr later, at $z\sim 6$.

After intense efforts to search for LyC emitters or LyC-leaking galaxies
(abbreviated to ``leakers''), several groups have identified such galaxies from low redshift up to $z \sim 4$ \citep[e.g.,][etc.]{Leitherer_2016, Izotov_2016a,Izotov_2016b, Izotov_2018a,Izotov_2018b, Vanzella_2016,Vanzella_2018,Vanzella_2020,DeBarros_2016,Steidel_2018,Fletcher_2019,Bian_2017,Wang_2019} with LyC escape fractions, \fesc, ranging from a few percent up to $\sim$ 72~\% in some cases \citep{Izotov_2018a, Vanzella_2020}. Although different methods and criteria have been used to search for these galaxies, some of the distinct observational features shared by many LyC emitters are strong \lya\ emission with a double-peaked \lya\ line profile \citep[see][]{Verhamme_2015,Verhamme_2017,Vanzella_2020} and intense optical emission lines with a high \oiii/\oii\ \footnote{In all the article we will use \oiii\ for [O~{\sc iii}] $\lambda$5007, \oii\ for [O~{\sc ii}] $\lambda\lambda$3726,3728, \oi\ for \Oi, \nii\ for [N~{\sc ii}] $\lambda$6584, \siii\ for \Siii\ and \sii\ for [S~{\sc ii}] $\lambda\lambda$6716,6731.} ratio \citep{Izotov_2018b, DeBarros_2016,Vanzella_2020}.
Following the suggestion by \cite{Jaskot_2013} and \cite{Nakajima_2014}  that high values of \oiii/\oii\ could pinpoint density-bounded galaxies, that is,\ LyC-emitting galaxies, this simple prediction was put to test by Izotov and collaborators, who selected compact star-forming galaxies with intense emission lines and \oiii/\oii$>5$, and who were thus able to find 11 LyC emitters at $z \sim 0.3-0.4$ with 100\% success rate using the COS spectrograph on-board the Hubble Space Telescope (HST).
The galaxies found in this way have properties comparable to the numerous galaxies at $z>6$. They are strong producers of ionizing photons, show significant LyC escape (\fesc $\sim 2-72$ \%), and are therefore good analogs of the sources of cosmic reionization \citep[cf.][]{Schaerer_2016}. 

Despite the fact that high \oiii/\oii\ was used as a basic selection criterion for these searches, it nevertheless turns out that \fesc\ does not well correlate with the \oiii/\oii\ ratio \citep[see][and discussions therein]{Izotov_2018b,Naidu_2018,Bassett_2019,Nakajima_2020}.
Similarly, \cite{Wang_2019} recently selected some galaxies by their \sii\  deficiency, which could indicate a density-bounded ISM, and were able to detect LyC emission from two low-$z$ galaxies with fairly low excitation (\oiii/\oii\ $\sim 1.3-1.5$). However, these authors were not able to find a correlation between the \sii\  deficiency and the LyC escape fraction of the known leakers. These findings are not entirely surprising, as it is well known that the \oiii/\oii\ ratio depends also on several other factors, including in particular the ionization parameter and  metallicity.  Whether or not density-bounded models are applicable to known LyC emitters therefore remains to be examined quantitatively.

On the other hand, the detailed analysis of UV absorption lines (including the Lyman series lines of hydrogen and other low-ionization lines) of the $z \sim 0.3-0.4$ leakers of Izotov and collaborators undertaken by \cite{Gazagnes_2018} and \cite{Chisholm_2018} revealed the presence of high-column-density gas that is optically thick to LyC radiation and co-exists with lines of sight, which allows the escape of UV ionizing and nonionizing radiation from these galaxies. This implies that the ISM of these galaxies cannot be described by simple, one-zone photoionization models, and that a more complex ISM structure should be accounted for to consistently describe these objects.
Therefore, testing one-zone photoionization models tailored to the properties (metallicities etc.) of the LyC emitters is required, as well as the  exploration of more complex structures to consistently describe their emission line spectra. These are two of the main goals of our study.

Specifically, we would like to know whether or not simple---that is,\ single--- 1D photoionization models can reproduce the observed emission-line properties of the LyC emitters, and if so, how well they do this. Comparing grids of photoionization models with observational diagrams of approximately $ 800$ low- to moderate-metallicity star-forming galaxies, \cite{Stasinska_2015} concluded that the strong lines indicate no strong LyC leakage (i.e.,\ $\fesc \la 10$\%) in galaxies with high \oiii/\oii, because the observed emission line ratios are not compatible with density-bounded H~{\sc ii} regions and therefore with a significant ionizing photon escape.
The same authors also stressed the importance of considering high- and low-excitation lines. Furthermore, they conjectured that some leakage might occur because of a nebular covering factor smaller than unity, and that \oi\ emission could come from optically thick clumps embedded in a density-bounded medium \citep{Stasinska_2015}.
Related to this, \cite{Plat_2019}  recently noted that low-$z$ LyC emitters show a higher \oi/\oiii\ ratio than their ionization-bounded photoionization models, in contrast to expectations from density-bounded models where \oi/\oiii\ should be reduced. To solve this discrepancy these latter authors invoke a contribution of hard radiation from active galactic nuclei (AGNs) or radiative shocks. However, no consistent model allowing for the observed leakage and escape fraction, including the geometrical constraints derived from the UV lines (as mentioned\ above), and explaining the intensities of the major emission lines has yet been made.

To progress further on these issues, here we compute both one- and two-zone photoionization models with metallicities tailored to those of the observed LyC emitters to quantitatively examine how such models compare to the various detailed observational constraints, including measured LyC escape fractions.  First, we re-examine the most important observed emission line ratios of the $z \sim 0.3-0.4$ leakers, and compare them to a large sample of star-forming galaxies of comparable metallicity, serving as a meaningful reference sample, which is presumably dominated by nonleaking galaxies. We then compare the data to standard one-zone photoionization models to see the successes and failures of these models. Finally, because one-zone models cannot properly reproduce the data, we are led to construct two-zone models which show a lot of interesting features and provide useful insight into the ISM and emission-line properties of the LyC emitters. 

If successful, consistent models should also allow us to better understand the origin of the different optical emission lines, their dependency on geometrical factors, leaking, and other factors. Hopefully, these models will lead to the development of a new diagnostic of LyC leakage using rest-frame optical emission lines that is applicable to other observations, especially of high-$z$ galaxies, for which many spectra will be obtained in the near future with the James Webb Space Telescope (JWST).

The need for multi-zone (or multi-component) models is not only of importance for LyC emitters. Indeed, numerous papers have pointed out the necessity to account for variations in density, ionization parameter, or geometry in order to simultaneously explain the observed strong emission line ratios of galaxies including different elements and ionization stages \citep[see][]{Stasinska_1996,Stasinska_2006}. Similarly, detailed studies of nearby galaxies with a large set of observations have clearly shown the limitations of one-zone models and the need for more complex and realistic representations \citep[e.g.,][]{Stasinska_1999,Cormier_2012,Cormier_2019,Lebouteiller_2017}. Here we explore some of these aspects with a focus on LyC-emitting galaxies.

This paper is organized as follows. 
In Sect.\ \ref{s_obs} we present our observational samples.
Section\ \ref{s_classical_mod} shows the main optical emission line diagrams used to compare to "classical" single-component photoionization models (both density- and ionization-bounded). We then construct two-component photoionization models including LyC leaking emission and compare them with observations (Sect.\ \ref{s_composite_models}).
Our results are discussed in Sect.\ \ref{s_discuss}.
Section \ref{s_conclude} summarizes and concludes the paper.
Throughout this paper, we assume a solar chemical composition following \cite{Asplund_2009} where $\log (Z/Z_{\odot})=12+\log({\rm O}/{\rm H})-8.69$.

% % % % % % % % % % % % % % % % % % % % % % %
\begin{figure}[htb]{
   \centering
   \includegraphics[width=9cm]{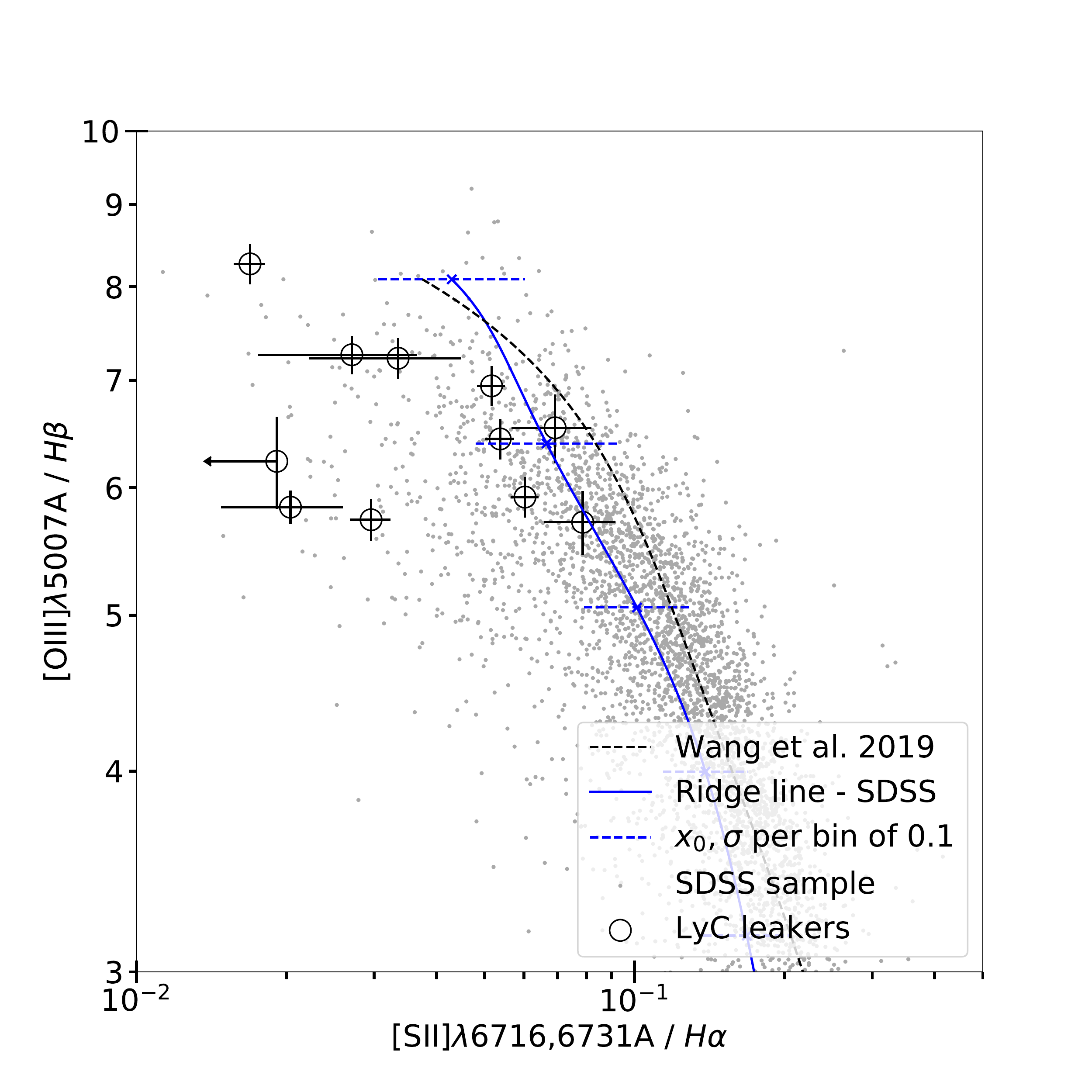}
   \caption{O3H$\beta$-S2H$\alpha$ diagram of the LyC leakers (open circles) and comparison sample with metallicities $\oh = 7.6-8.2$ (gray dots). The plain blue line shows the ridge line of our SDSS sample, and the black dashed line is the ridge line from \cite{Wang_2019} extrapolated to high \oiii/\hb\ values. Many leakers, especially the strongest ones, show a deficiency in \sii\ with respect to this latter ridge line.}
   \label{fig_s2}}
\end{figure}

% % % % % % % % % % % % % % % % % % % % % % %

\section{Optical emission line properties of LyC emitters} 
\label{s_obs}

In the present work we study eleven $z\sim 0.3-0.4$ Lyman continuum emitters discovered with the COS spectrograph on-board HST and whose properties have been amply discussed in several papers \citep{Izotov_2016a, Izotov_2016b, Izotov_2018a, Izotov_2018b,Schaerer_2016, Verhamme_2017, Gazagnes_2018, Chisholm_2018,Schaerer_2018}.
The metallicities of these galaxies are in the range $\oh = 7.65-8.16$, with a median oxygen abundance of $\oh$ of 7.91.

These sources represent currently the best sample of local LyC emitters, for which a wide range of observations is available, including LyC, UV, and optical spectroscopy. Furthermore, \cite{Gazagnes_2018} and \cite{Gazagnes_2020} provided a detailed analysis of their UV absorption lines (including Lyman series lines and other low-ionization absorption lines), which provides important geometrical constraints on the ISM of these galaxies. Here we analyze the optical emission line properties of these LyC emitters.

Before discussing the observed emission line properties of the LyC emitters, we briefly summarize the observational data used, including the main galaxies of interest (11 LyC emitters) and an appropriate comparison sample.

\subsection{Observational data}

\subsubsection{Lyman continuum emitters}

The emission line measurements from the SDSS spectra of the LyC emitters are reported in the discovery papers 
\citep{Izotov_2016a, Izotov_2016b, Izotov_2018a, Izotov_2018b}.
We recently obtained new VLT spectra of 5 of the 11 leakers (J0925+1403, J0901+2119, J1011+1947, J1154+2443 and J1442-0209) 
with the XShooter spectrograph. We use the improved emission line measurements from these spectra,
and in particular also the \Siii\ measurements, which are present in the near-IR spectra of these galaxies.
A more detailed description of these observations and results are presented in \cite{Guseva_2020}.
All emission line ratios reported here refer to extinction-corrected ratios.

\subsubsection{Comparison sample}
For comparison we use a sample of compact star-forming galaxies from the Sloan Digital Sky Survey (SDSS) Data Release 14  \citep{Abolfathi_2018} analyzed in earlier publications \citep{Guseva_2019}, and for which the [O~{\sc iii}] $\lambda$4363 line is detected with an accuracy better than $4 \sigma$, allowing direct abundance 
determinations using the $T_e$ method. In practice, from the list of 5607 galaxies with direct metallicity measurements, we retain 4571 galaxies in the metallicity range $\oh = 7.6-8.2$, for comparison with the LyC emitters discussed here.

\begin{figure}[tb]{
   \centering
   \includegraphics[width=9cm]{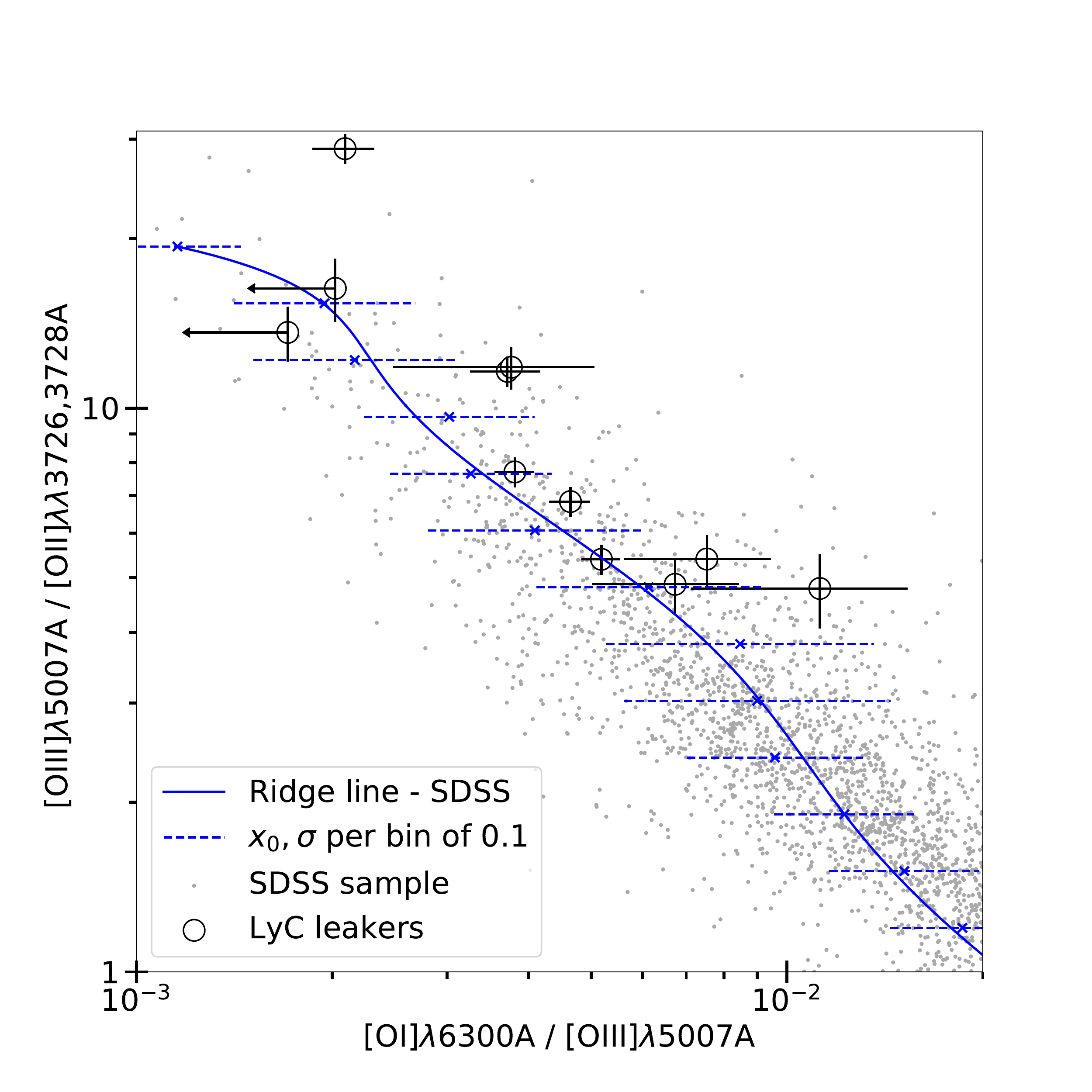}
   \caption{O32-O13 diagram.}
   \label{fig_o321}}
\end{figure}

\subsection{Optical emission line properties of LyC emitters} 
\label{obs_trends}
The emission line properties of the LyC emitters and the comparison sample are shown in Figs.\ \ref{fig_s2},
\ref{fig_o321}, and \ref{fig_BPT}.
By selection, all the galaxies studied here, including the LyC emitters, have emission lines ratios 
compatible with those of star-forming galaxies, that is,\ they fall in the range of stellar photoionization sources in the 
classical diagrams by \cite{Veilleux_1987} used for spectral classification \citep[including the \nii\ BPT diagram introduced by][]{1981PASP...93....5B}. 
Also, by selection, the sources studied here have a relatively low metallicity, which
implies a fairly high excitation, reflected for example\ by a high \oiii / \hb\ ratio.
However, at a finer level, the LyC emitters show some interesting distinctive properties, which we now discuss.

\subsubsection{A \sii\  deficiency in LyC emitters}

In Fig.\ \ref{fig_s2} we show a classical diagnostic (O3H$\beta$-S2H$\alpha$) involving the \sii/\ha\ ratio,
which was used by \cite{Wang_2019} to select LyC-emitter candidates among more metal-rich and therefore
lower excitation (lower \oiii/\hb) sources.
Clearly several, if not a majority, of the eleven confirmed leakers show fairly weak \Sii\ emission, or  \sii/\ha\ ratios close to or below the typical value for galaxies from our comparison sample at the same \oiiil/\hb\ ratio.
To quantify this comparison we derive an SDSS ``ridge line'' which marks the median location of normal star-forming galaxies 
in the same range of O/H (metallicity) as our LyC emitters. This was obtained by binning the data in logarithmic bins of 0.1 dex along the \oiii/\hb\ axis with each bin having at least ten galaxies. 
Each bin was then fitted by a Gaussian distribution of a center $x_{0}$ and a standard deviation $\sigma$. 
The ridge line defined by the $x_{0}$ was then fitted with an eight-order polynomial, 
\begin{equation}
    \centering
    \log x= \sum_{n=0}^{8} \alpha_{n} (\log y)^{n},
    \label{eq_ridge}
\end{equation} 
where $y=$\oiiil/\hb, and the coefficients $\alpha_{n}$ are reported in Table \ref{tab_coeffs}. Table \ref{tab_coeffs2} presents the coefficients of the ridge lines derived in other diagnostic plots $x$ vs $y=$\oiii/\oii.

\begin{table}[tb]
\centering
\caption{Coefficients of the ridge lines in classical diagrams (Figs. \ref{fig_s2}, \ref{fig_BPT} and \ref{fig_BPT_combined}), determined for the entire galaxy sample and described by Eq.\ \ref{eq_ridge}. Here $y$=\oiiil/\hb.}
\label{tab_coeffs}
\begin{tabular}{rrrrr}
    %\hline
    $x$ & \nii/\ha & \sii/\ha & \oi/\ha \\
    \hline
    $\alpha_{0}$ & -2.40 & -2.00 & -3.27 \\
    $\alpha_{1}$ & 5.22e1 & 3.78e1 & 4.33e1 \\
    $\alpha_{2}$ & -5.74e2 & -3.98e2 & -4.33e2 \\
    $\alpha_{3}$ & 3.14e3 & 2.11e3 & 2.24e3 \\
    $\alpha_{4}$ & -9.77e4 & -6.40e3 & -6.68e3 \\
    $\alpha_{5}$ & 1.80e4 & 1.16e4 & 1.20e4 \\
    $\alpha_{6}$ & -1.94e4 & -1.23e4 & -1.28e4 \\
    $\alpha_{7}$ & 1.14e4 & 7.16e3 & 7.48e3 \\
    $\alpha_{8}$ & -2.79e3 & -1.74e3 & -1.83e3\\
    \hline
\end{tabular}
\end{table}

\begin{table}[tb]
\centering
\caption{Coefficients of the ridge lines in oxygen and sulfur diagrams (Figs. \ref{fig_o321},\ref{fig_o_s} and \ref{fig_OS_combined}), described by Eq.\ \ref{eq_ridge}. Here $y=$\oiii/\oii.}
\label{tab_coeffs2}
\begin{tabular}{rrrr}
    %\hline
    $x$ & \oi/\oiii & \sii/\siii \\
    \hline
    $\alpha_{0}$ & -1.67 & 4.13e-1 \\
    $\alpha_{1}$ & -1.07 & -5.15e-1  \\
    $\alpha_{2}$ & 4.68e-1 & -1.54  \\
    $\alpha_{3}$ & 1.46 & 1.16e1 \\
    $\alpha_{4}$ & -9.67e-1 & 1.30 \\
    $\alpha_{5}$ & -5.33 & -9.20e1 \\
    $\alpha_{6}$ & 4.36 & 1.75e2  \\
    $\alpha_{7}$ & 2.01 & -1.28e2  \\
    $\alpha_{8}$ & -1.84 & 3.31e1\\
    \hline
\end{tabular}
\end{table}

\begin{figure}[htb]{
   \centering
   \includegraphics[width=9cm]{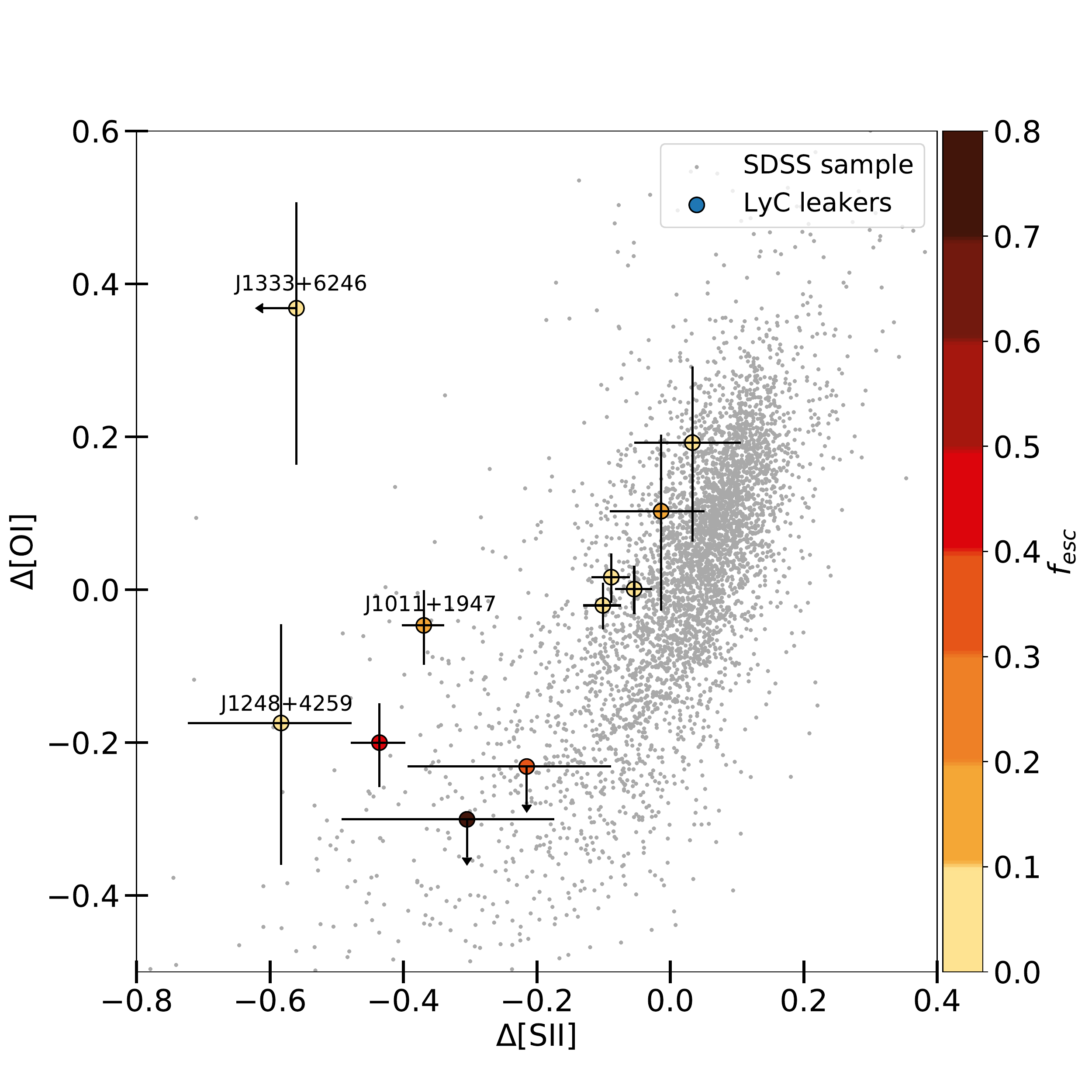}
   \caption{$\Delta$\oi\ vs. $\Delta$\sii\ derived from classical diagrams O3H$\beta$-O1H$\alpha$ and O3H$\beta$-S2H$\alpha$. The three strongest leakers (J1256+4509, J1154+2443 and J1243+4646, by increasing order of $f_{esc}$) are reprented by orange to brown circles in the lower part of the plot. The other galaxies discussed in this paper are labeled. The LyC emitters show a clear displacement in this plane compared to the comparison sample. }
   \label{fig_deltas_obs}}
\end{figure}

For subsequent discussions, we also calculate $\Delta$\sii, the lateral shift from the ridge line,
\begin{equation}
    \Delta[{\rm SII}] = \log\left([{\rm SII}]/\ha \right)- f\left(\log\left([{\rm OIII}]/ \hb\right)\right),
    \label{e12_bis}
\end{equation}
to quantify deviations of individual galaxies from the average observed line ratio of the entire comparison sample.
With this definition $\Delta$\sii <0 indicates a deficit in \sii/\ha\ with respect to the average at the same \oiiil/\hb\ ratio. 
We note that the sulfur abundance ratio (S/O) is constant and shows a fairly small scatter over a wide range of metallicities, even beyond the range of O/H considered here \citep[see, e.g.,][]{Guseva_2020}. Therefore, the mean trends described by this ridge line should be quite independent of the exact S/O abundance. 

As shown in Fig.\ \ref{fig_s2}, five of our LyC emitters are within 1~$\sigma$ of the ridge line, whereas the rest are systematically offset to lower \sii/\ha\ ratios, that is,\ they show an \sii\  deficit, an empirical finding already pointed out by \cite{Wang_2019}. Several reasons could be invoked to explain such a behavior: S/H abundance variations; a decrease of \sii\ due to the loss of the outer, neutral regions related to the escape of LyC radiation; a lower contribution of diffuse ionized gas (DIG), which is in general known to ``boost'' \sii\ emission \citep[cf.][]{Sanders_2020}; or possibly other explanations.
The \sii\ deficit and its link with the observed LyC escape fraction are discussed below, providing a  more quantitative description of the relationship (Sect.\ \ref{s_links_fesc}).

\subsubsection{An excess of [OI]/[OIII]}
Another peculiarity of the LyC emitters is found in the diagram comparing the three ionization stages of oxygen traced by the optical line ratios \oiii/\oii\ and 
\oi/\oiii, as shown in Fig.\ \ref{fig_o321} (denoted  the O32-O13 diagram hereafter).
Indeed, naively one would expect to find a deficit of \oi\ emission in LyC emitters as they should be highly ionized and possibly lacking neutral regions where \oi\ is emitted. This is also shown by simple density-bounded photoionization models \citep[see Figs.\ \ref{fig_o_s} and \ref{fig_o32_fesc_BOND}, and][]{Stasinska_2015}. However, as also noted by \cite{Plat_2019}, \oi\ is detected in 9 out of the 11 leakers of this study, and the observed \oi/\oiii\ ratios of the LyC emitters are surprisingly high, shifted  to even higher than average values when compared to the ridge line defined by the comparison sample in Fig.\ \ref{fig_o321}. This empirical finding clearly shows the presence of neutral oxygen in the LyC leakers, which already indicates that the ISM in these galaxies can probably not be described by simple density-bounded models, and whose presence and intensity needs to be explained quantitatively by models.  

\subsubsection{A shift in low-ionization line properties} 

In Fig.\ \ref{fig_deltas_obs}, we show the displacement relative to the ridge lines in O3H$\beta$-O1H$\alpha$ and O3H$\beta$-S2H$\alpha$ diagrams (see columns 2 and 3 in Table \ref{tab_coeffs}). In this plane, we see a clear shift from the SDSS comparison sample of presumably nonleaking galaxies. Interestingly, the three most leaking galaxies (the three reddest circles, $f_{esc} \ga 38\%$) in our sample stand in the lower-left corner with low $\Delta$\sii\ and $\Delta$\oi\ (< -0.2 dex). On the other hand, most of the small leakers in our sample (yellow circles with $f_{esc}$ \la 10\%) exhibit a moderate deficiency in \sii\ (-0.2 < $\Delta$\sii < 0) and \oi\ excesses ($\Delta$\oi\ \ga 0). However, we note that three galaxies in our sample with a large deficit in \sii\ ($\Delta$\sii \la -0.3) have intermediate (J1011+1947, $f_{esc}$=11,4\%) to low escape fractions (J1248+4259, J1333+6246, labeled on the plot). One of them (J1333+6246) strikingly stands apart with a nondetection of the \sii\ line but the largest \oi\ excess in the sample. A more detailed analysis of these shifts in \oi\ and \sii, including a comparison to models, is given in Sect.\ \ref{s_links_fesc}.

\subsubsection{Other emission-line peculiarities of LyC emitters}

In the \nii\ BPT diagram (O3H$\beta$-N2H$\alpha$, Figs. \ref{fig_BPT} and \ref{fig_BPT_combined}) it is possible to note a slight offset of most of the leakers with respect to the average of the comparison sample. Likely related to this, \cite{Guseva_2020} noted that the N/O abundance of LyC emitters is again higher than the average of star-forming galaxies at the same O/H abundance. The origin of this behavior is currently unknown. 

In another diagram discussed further below, relating \oiii/\oii\ to \Sii/\Siii\ (Fig.\ \ref{fig_o_s}), we note that four out of five LyC emitters, for which we have near-IR measurements of \Siii, appear also shifted from the average \Sii/\Siii\ ratio in a similar fashion as for \oi/\Oiii. More measurements will show if this behavior is systematic for LyC leakers or not. 

Finally, \cite{Izotov_2017}  proposed that LyC-emitting galaxies show peculiar line ratios of weak He~{\sc i} lines. As shown by \cite{Guseva_2020}, our Xshooter observations confirm the trend suggested by this study.

To gain further insight into the observed behavior of the emission line ratios we now proceed to comparisons with photoionization models, first with simple and then more complex geometries.

%%%%%%%%%%%%%%%%%%%%%%%%%%%%%%%%%%%%%%%%%%%%%%%%%%%%%%%%%%%%%%%%%%%%%%
\section{Comparison to classical photoionization models}
\label{s_classical_mod}

The first natural step in a comparison between the observed emission line properties of LyC emitters is to examine how simple 1D photoionization models allowing for varying degrees of ionizing photon escape behave and compare with the data.  
% % % % % % % % % % % % % % % % % % % % % % %
\subsection{The models: BOND models from the 3MdB database} \label{BOND_description}

To examine the power of simple 1D photoionization models, we used Cloudy photoionization models assembled in the BOND (Bayesian Oxygen and Nitrogen abundance Determinations) grid of models \citep{ValeAsari_2016}, which is available on the Mexican Million Models database \citep[3MdB,][]{Morisset_2015}. This grid was designed to be used with the BOND code, a Bayesian code, which simultaneously derives oxygen and nitrogen abundances in giant H~{\sc ii} regions. One of the specificities of the BOND code is that it is independent of any assumption on relations between O/H and N/O. Therefore, the grid spans a wide range in O/H, N/O, and ionization parameter U, and covers different starburst ages and nebular geometries (spherical bubble or thin shell). The spectral energy distribution (SED) of the ionizing radiation is obtained from the population synthesis code PopStar \citep{Molla_2009} for a Chabrier \citep{Chabrier_2003} stellar initial mass function (IMF) between 0.15 and 100 \msun. Varying starburst ages describe variations in the ionizing radiation field hardness, which arise due to the evolution of the stellar populations ionizing the H~{\sc ii} regions.
A subset of the BOND model grid has been used by \cite{Stasinska_2015}, who  also studied the effect of LyC leakage. 
In the current study, we use models from the 3MdB\_17 database\footnote{The 3MdB website describes the 3MdB and 3MdB\_17 databases at {\tt https://sites.google.com/site/mexicanmillionmodels/}} (ref='BOND\_2'), which is an updated version of 3MdB using Cloudy v17.02 \citep{Ferland_2017} and covering a somewhat refined grid. 

For our study, we used a sub-grid of BOND models, with parameters appropriate for the LyC-emitting galaxies and the comparison sample. Specifically, the metallicity range is limited to \oh=7.6-8.2 (with a step of 0.2), covering the observed metallicities of the LyC emitters. The ionization parameter $\log U$ varies between -4 and -1 (with a step of 0.25), the age between 1 and 3 Myr (in steps of 1 Myr). The models are spherical filled bubbles with a fixed density of $100$\ cm$^{-3}$, and abundance ratios of N/O= $10^{-1}$ and S/O=$10^{-1.66}$. The choice of a spherical filled bubble geometry influences the ionization structure of the modeled  H~{\sc ii} regions: based on \cite{Stasinska_2015}, we expect thin shell models to produce fainter emission from low-ionization-potential species (e.g., \oi, \sii) with respect to higher ionization potential species (e.g., \oiii, \siii). Specifically, this would induce a lateral shift to the left-hand side in the O32-O13 and O32-S23 diagnostics plots (Fig.\ \ref{fig_o_s} and \ref{fig_OS_combined}), which we examine in Section \ref{spectral_diag}. 

To describe density-bounded models, we used a parameter $H_{\beta}frac$ 
which cuts the nebula at different radii corresponding to a range of 10--100\% of the total \hb\ luminosity.\footnote{Since the study of \cite{ValeAsari_2016}, the grid in $H_{\beta}frac$ has been refined and is now available for steps of 10\%.} This parameter corresponds to the fraction of $\hb$ luminosity emitted by the entire modeled galaxy relative to the $\hb$ luminosity that would be emitted by the same model if it was ionization-bounded. Since the $\hb$ emissivity is roughly constant throughout the  H~{\sc ii} region, this quantifies the ratio of the radius of the truncated, density-bounded model to the Str\"omgren radius of its ionization-bounded analog. 
We directly translate $H_{\beta}frac$ into the corresponding LyC escape fraction by defining $f_{esc}(\hb)$ = 1 - $H_{\beta}frac$ which we use as a proxy to estimate how much leakage is allowed by a given density-bounded model.
We refer to the models used here as simple, one-zone models.

\subsection{The spectral diagnostics}
\label{spectral_diag}

We now compare the one-zone photoionization models to the observations using the classical diagrams introduced by \cite{Veilleux_1987} for spectral classification, the above-mentioned O32-O13, and other diagrams.

\begin{figure*}[htb]
{\centering
\includegraphics[width=18cm]{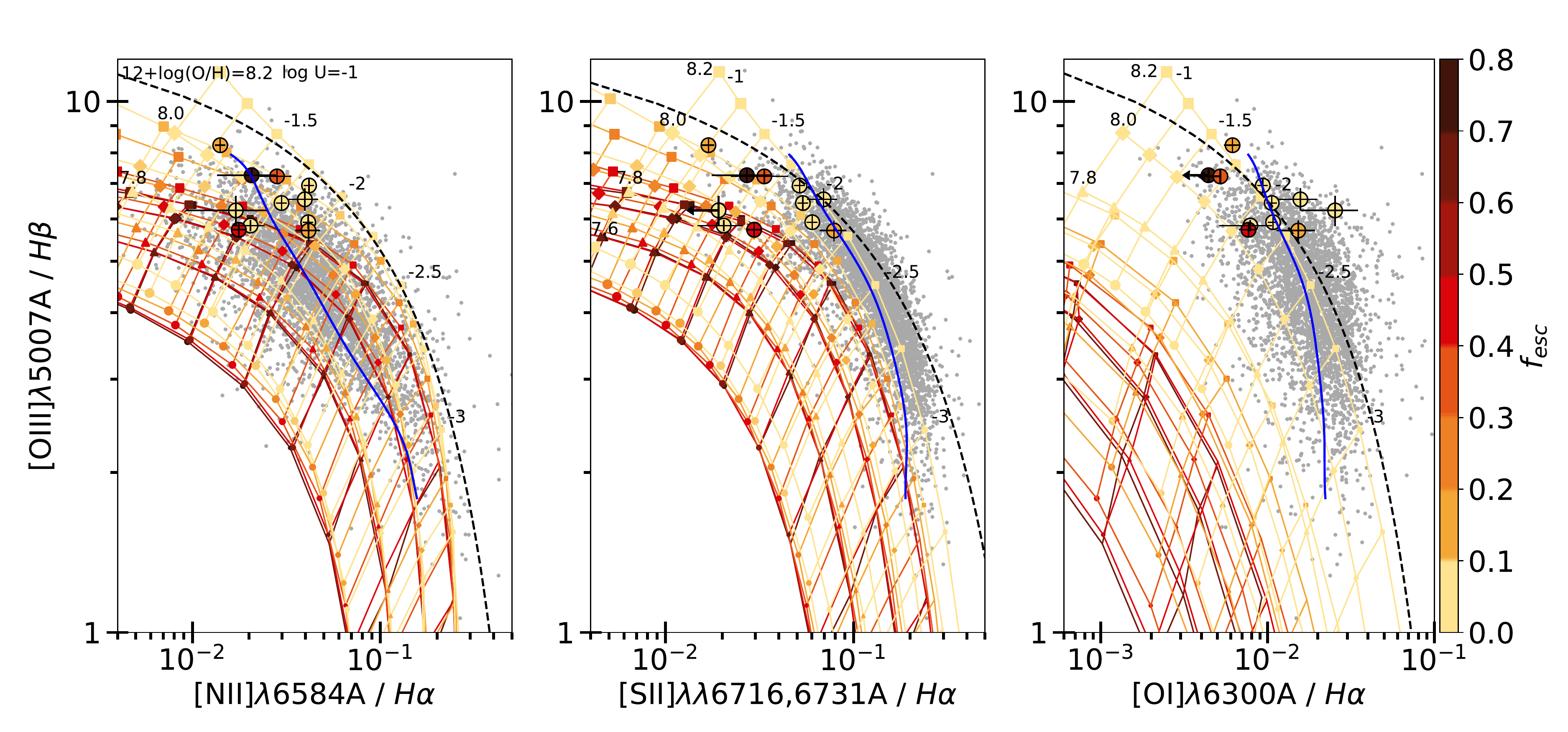}
\includegraphics[width=18cm]{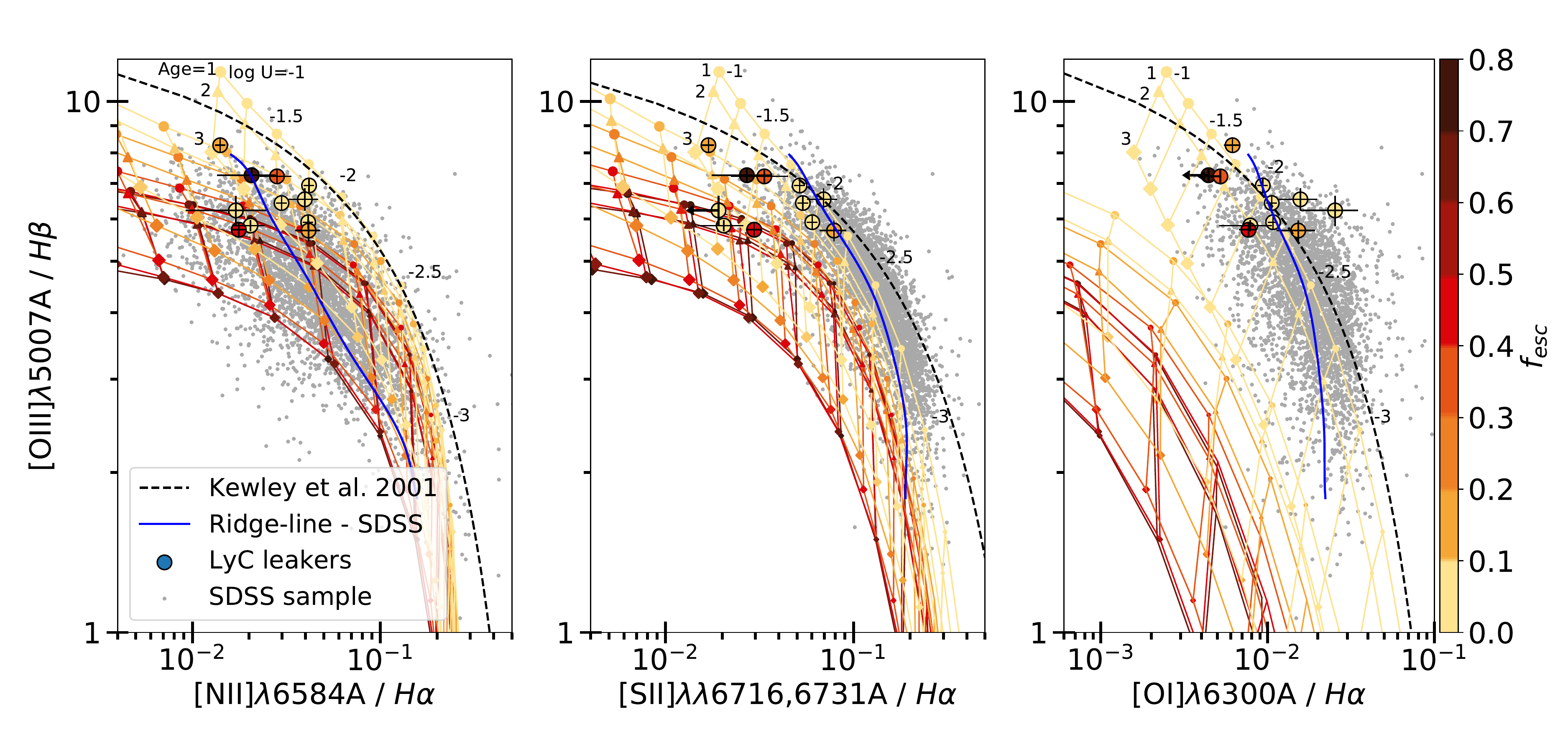}
\caption{Classical diagrams with BOND models and observations. Photoionization models are represented by the colored symbols: the colorbar represents the variation of the escape fraction and the size of symbol is proportional to $\log(U)$. The shape either represents the variation of metallicity (first row: for the age of 1 Myr) or stellar age (second row: for the metallicity of 8.2).
The observed data points (gray dots) show the SDSS galaxy sample in the restricted metallicity range, $\oh = 7.6-8.2$.
The plain blue lines show the ridge lines of the entire sample. Dashed lines correspond to the AGN delimitations from \cite{Kewley_2001}.}

\label{fig_BPT}
}
\end{figure*}

\subsubsection{Classical diagrams}

Figure \ref{fig_BPT} shows the LyC emitters, the comparison sample, and the one-zone models in the classical diagrams from \cite{Veilleux_1987}, including the \nii\ BPT diagram. As before, we also indicate the ridge line, describing the location of the average SDSS galaxies in the corresponding plots (cf.\ Sect.\  \ref{obs_trends}). For illustration, dashed lines indicate the theoretical delimitations between star-forming galaxies (below) and AGN (above), according to the models of \cite{Kewley_2001}.

Overall, the ionization-bounded ($\fesc=0$) photoionization models cover reasonably well the observed line ratios of the entire galaxy sample including the LyC emitters. The O3H$\beta$-N2H$\alpha$ diagram is better covered by these models than the
O3H$\beta$-S2H$\alpha$ and O3H$\beta$-O1H$\alpha$ diagrams. For example, the ridge line in the \sii\ diagram is located outside the parameter space examined here; discussing the origin of these offsets is beyond the present scope.
The position of the LyC leakers indicates high ionization parameters and young stellar ages of the ionizing populations; the latter is in good agreement with the young ages found from the observed UV spectral features and from spectral modeling of these galaxies \citep[see, e.g.,][]{Izotov_2019}.

The second most important conclusion from Fig.\ \ref{fig_BPT} is that the domain of the emission line ratios observed in the LyC emitters is only covered by ionization-bounded photoionization models, that is,\ models with $\fesc =0$ or clearly only for low values of $\fesc < 0.1$, which is in contradiction with the observed escape fractions reaching up to $\sim$ 70\% for some of the leakers.
As already shown by \cite{Stasinska_2015}, increasing the escape fraction in these simple, one-zone 1D photoionization models strongly decreases the \sii\ and \oi\ emission, which results from trimming the outer parts of the  H~{\sc ii} region. This shows that overall the observed LyC emitters (or at least those with a significant LyC escape fraction) cannot be reproduced by simple density-bounded models, as these systematically underpredict lines of species with low ionization potentials, such as \oi\ and \sii.

\begin{figure*}[htb]
{\centering
\includegraphics[width=15cm]{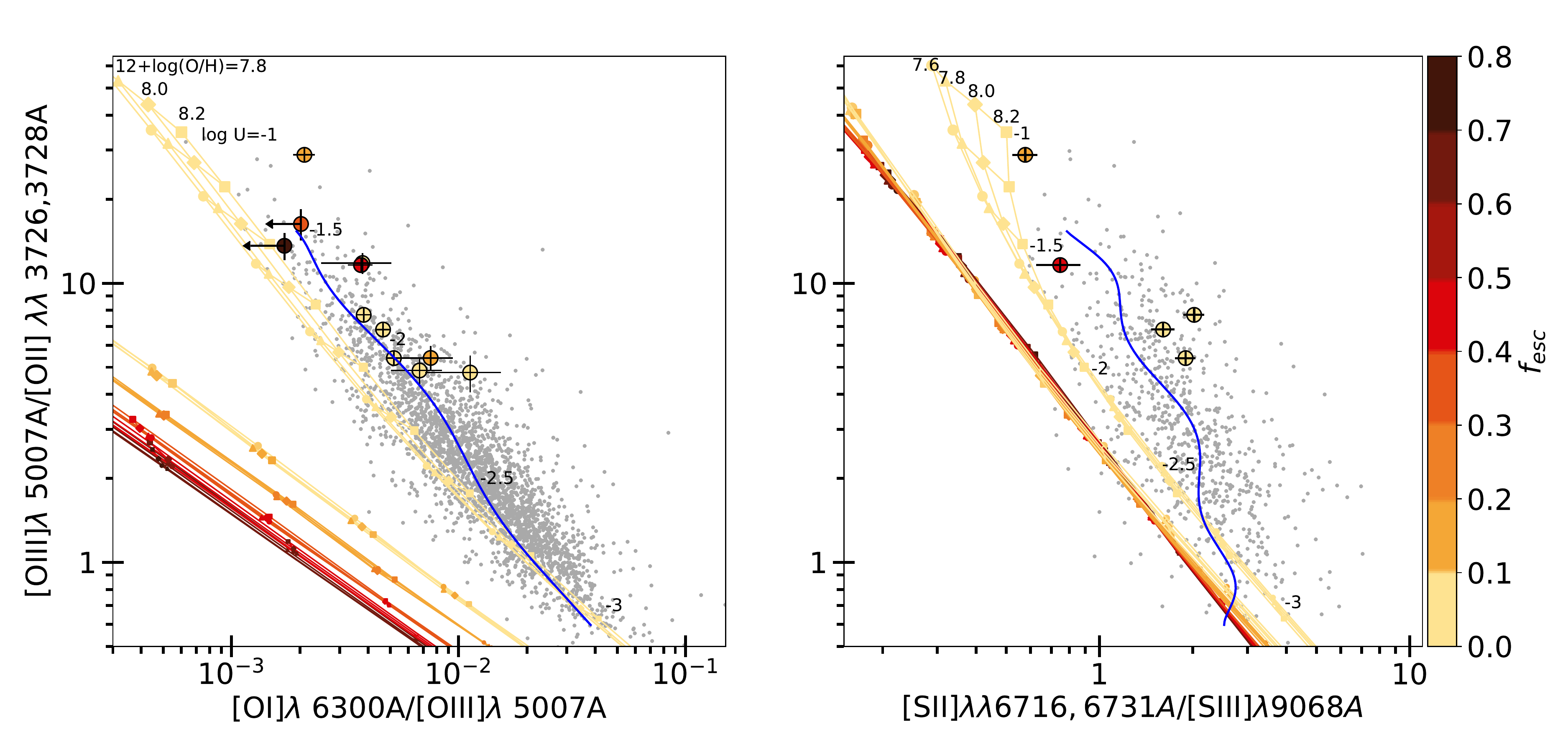}
\includegraphics[width=15cm]{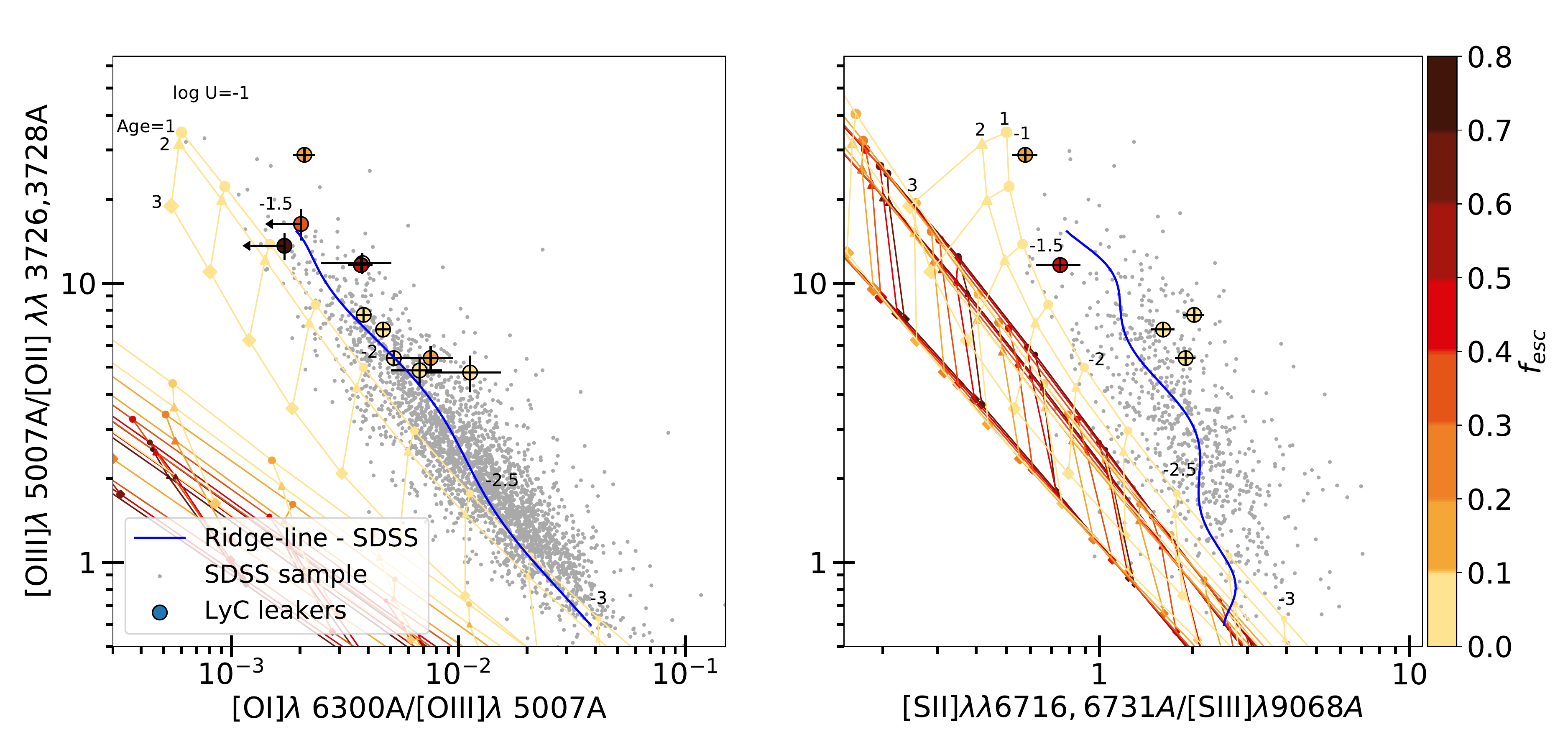}
\caption{O32-O13 and O32-S23 diagnostics with BOND models. Photoionization models are represented by the colored symbols: the colorbar represents the variation of the escape fraction and the size of symbols is proportional to $\log(U)$. 
The shapes of model dependencies reflect the variations either with metallicity or stellar age and are shown and labeled in the top and bottom panels, respectively.}
\label{fig_o_s}
}
\end{figure*}

\begin{figure*}[htb]
{\centering
\includegraphics[width=15cm]{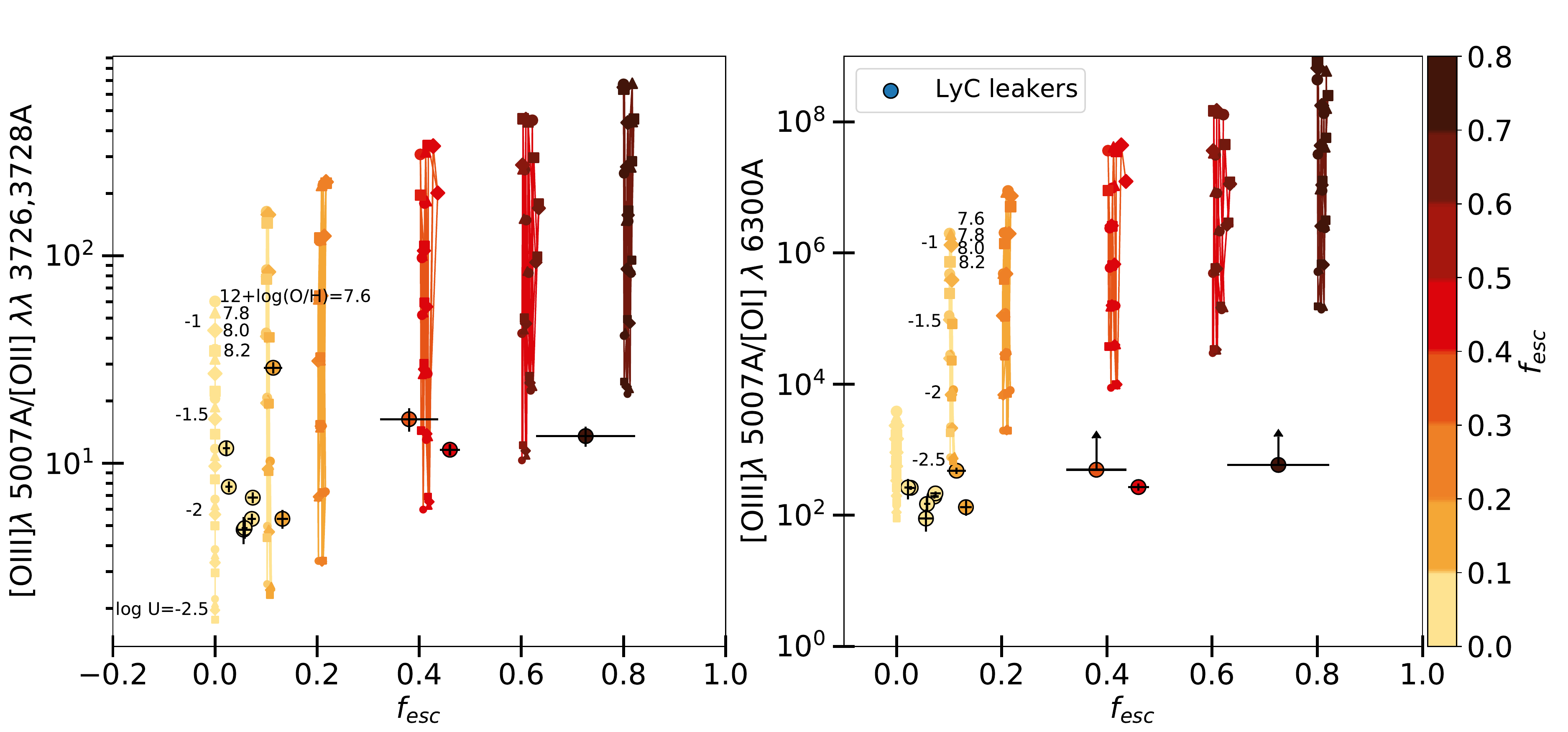}
\caption{Observed and predicted \oiii/\oii\ (left) and \oiii/\oi\ (right) line ratios as a function of the LyC escape fraction, comparing the observed LyC emitters and the one-zone photoionization models with $\log(U) \protect\ga -2.5$. We note the very high range of the predicted 
\oiii/\oi\ ratio.}
\label{fig_o32_fesc_BOND}
}
\end{figure*}

\subsubsection{Diagnostics of the low-ionization regions}

In Fig.\ \ref{fig_o_s} we examine complementary and probably more powerful emission line diagnostics to probe the ionization structure and therefore the different zones of the ionized ISM in the LyC emitters. By focusing on different ionization stages of the same element, we also avoid dependencies on elemental abundance ratios and metallicity, at least to first order. These O32-O13 and O32-S23 diagnostics highlight two main points that motivate the introduction of more complex models in the Sect. \ref{s_composite_models}:

(1) Although the oxygen line ratios (O32-O13) predicted by simple ionization-bounded models reproduce the SDSS sequence with ionizing spectra of young populations ($\sim 1-2$ Myr) quite well, the models are not able to account for galaxies with a certain \Oi\ excess, that is,\ a  
higher-than-average \oi/\oiii\ ratio at a given \oiii/\oii\ ratio, which is observed for the LyC emitters and other sources (see Fig.\ \ref{fig_o_s} left). For \sii/\siii, shown on the right panel of this figure, the deficiency of the models is even more striking, as they fail to reproduce the bulk of the observations. The two discrepancies might arise from the fact that the models do not take into account a DIG component, which can significantly contribute to the emission of \oi\ and \sii\ \citep{Sanders_2017,Sanders_2020}. 
Alternatively, these discrepancies could also be attributed to the presence of shocks, which are not included in our models \citep[see, e.g.,][]{Plat_2019} because of problems with atomic data for sulfur \citep{Izotov_2006, Kewley_2002, Kewley_2019}, or others. The sulfur line ratio \sii/\siii\ is also sensitive to more subtle effects including stellar winds, outflows, and photoevaporation of photo-dissociation regions (PDRs) in Galactic dense gaseous environments \citep[e.g.,][]{Westmoquette_2013, McLeod_2015}.

(2) The density-bounded models allowing for LyC photon escape do not at all reproduce the observed oxygen and sulfur line ratios of the leakers, as already noted above. On the contrary, while leakers populate the upper part of the SDSS sequence, the density-bounded models produce very low emission in \oi\ and \sii, which drives them in the opposite direction to the bottom-left corner of our diagnostic plots. Figure \ref{fig_o32_fesc_BOND} represents the \oiii/\oii\ and \oiii/\oi\ ratios predicted by the models as a function of their escape fraction. This diagram emphasizes the discrepancy between \oiii/\oi\ and \fesc, showing that \oiii/\oi\ is overestimated by several orders of magnitude. This clearly shows that moderate to high escape fractions cannot be produced by simple density-bounded one-zone models without significantly decreasing the weak line emission of \oi\ and \sii.

In short, although density-bounded models would explain the Lyman-continuum-escaping photons, they clearly do not match the spectral signatures arising from the outer part of the  H~{\sc ii} region and observed in LyC emitters. On the other hand, while ionization-bounded models can describe the overall properties of most galaxies (the comparison sample)  fairly well---except for sulfur line ratios---, they underpredict the emission from low-ionization species (e.g.,\ the \oi\ line) in the LyC emitters, and most importantly, they do not account for the escape of LyC photons. These failures of simple, one-zone photoionization models clearly indicate that more complex models are required to describe the observed emission line properties of the LyC emitters. We therefore examined composite photoionization models based on the combination of two distinct zones with different properties.

\section{Composite "two-zone" photoionization models}
\label{s_composite_models}

\subsection{Motivation and method}

As demonstrated above (Sect.\ \ref{spectral_diag}), single-component (one-zone) photoionization models fail to simultaneously reproduce the observed line ratios and escape fractions of LyC leakers. To answer our problem (1) from Sect. 3.2.2 (insufficient emission in low-ionization lines), we introduce a component of low ionization (low $U$) that would mimic a DIG contribution in star-forming galaxies. The relative contribution of this component is described by a factor $\omega$ that varies from one galaxy to another. In the case of LyC emitters, although they are expected to be relatively poor in DIG \citep[see][]{Oey_2007}, this component could effectively represent the effect of low-ionization channels carved into the ISM through which some LyC photons escape, and which may contain gas with a low-volume filling factor, such as clumps or filaments. As an answer to problem (2) from Sect. 3.2.2 (inconsistency between the escape fraction and the presence of low-ionization lines), we include in our two-zone model the possibility for LyC photon escape from one region, while the second remains ionization-bounded. 

A two-zone, ``picket-fence''-type model, was also suggested based on other observations.
 \cite{Gazagnes_2018}, \cite{Chisholm_2018}, and \cite{Gazagnes_2020}  analyzed the UV absorption lines of hydrogen and low-ionization metal lines of the LyC emitters studied here. Their work clearly shows the presence of absorption lines formed in the neutral gas, with column densities that are too high for the escape of LyC photons. Furthermore, the depths of the saturated absorption lines show that channels of low column density must be present in these galaxies, allowing LyC escape. For our purposes, their model is basically equivalent to our two-zone model.
As we show below, such an idealized two-zone model can also capture and describe the observed optical emission lines of the LyC emitters.

\begin{figure}[tb]
{\centering
\includegraphics[width=9cm]{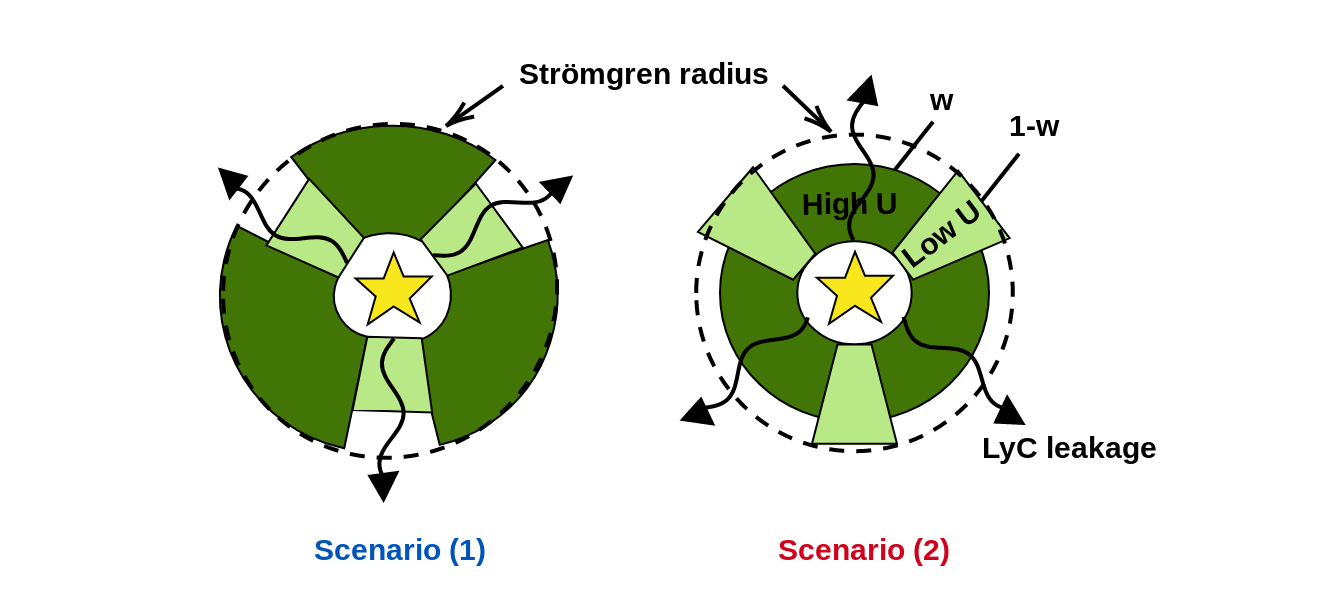}
\caption{Schematic illustration of our two-zone photoionization models for LyC emitters, combining an ionization region with density-bounded regions chosen to have different ionization parameters (``high'' and ``low'' $U$). In the "density-bounded channels"  scenario, ionizing photons escape from the low-$U$ region, while in the "density-bounded galaxy" scenario, they escape from the high-$U$ region with $f_{esc}$=0 for the nonleaking region and $f_{esc}$ between 0\% and 80\% for the leaking region.
}
\label{fig_scenario}}
\end{figure}

In practice, we have considered two different scenarios that are represented in Fig.\ \ref{fig_scenario}.
Each scenario is based on a linear combination of two models from the BOND database (see Sect.\ \ref{BOND_description}), one with a high-ionization parameter weighted by a factor $\omega$, and one with a low $U$ weighted by a factor $(1-\omega)$. In each scenario, we allow only one component to be leaking.

The high-$U$ component corresponds to the main body of the galaxy which is highly ionized and homogeneous (filling factor $\epsilon \sim 1$). We briefly discuss here the motivation for introducing a low-ionization component. The BOND models are parameterized by an input volume-averaged ionization parameter that, for a given geometry, is proportional to the ionization parameter of the corresponding ionization-bounded model at the Str\"omgren radius \citep[see Equation 4 in][]{Stasinska_2015}. For ionization-bounded models, the ionization parameter at the Str\"omgren radius is defined by the following equations:

\begin{equation}
    U(r = R_{S}) = \frac{Q}{4\pi R_{S}^{2}nc},\,\, {\rm and} \,\,
    R_{S} = \left(\frac{3Q}{4\pi\alpha(T_e)n^{2}\epsilon}\right)^{1/3},
    \label{e3}
\end{equation}
where $Q$ is the number of ionizing photons, $n$ the density of hydrogen, $\alpha(T_e)$ the recombination coefficient of hydrogen, and $\epsilon$ the filling-factor for an inhomogeneous medium ($0 \la \epsilon \la 1$).
Equation (\ref{e3}) leads to: 
%$$\rm{U = A (Qn\epsilon^{2})^{1/3}}$,
\begin{equation}
    U(r = R_{S}) = A (Q n \epsilon^{2})^{1/3},
    \label{e34}
\end{equation}
where $A$ is almost constant ($A \approx 1.3 \times 10^{-21}$cm s$^{1/3}$ for $T_{e}=10^4$ K), which means that both low-density and clumpy mediums ($\epsilon \ll 1$) will correspond to a model with a low input ionization parameter. We note that the density-bounded models created by imposing a cut in radius to the full ionization-bounded model do not necessarily have a mean ionization parameter that matches their input ionization parameter. The two components do not have to share exactly the same Str\"omgren radius but one can ensure that the physical sizes are compatible by adjusting the filling factor of one component to match the size of the other. We now further describe the scenarios in Fig.\ \ref{fig_scenario}.

{\em Density-bounded channels (DBC)}: Ionization-bounded galaxy with density-bounded channels. The galaxy is ionization-bounded but LyC photons can escape through channels, which are matter-bounded.
Such holes could be carved in the ISM by supernovae explosions or could result from hydrodynamical fragmentation effects induced by turbulence or stellar feedback \citep[e.g.,][]{Kakiichi_2019,Hogarth_2020}. We expect such mechanisms to efficiently clear out channels, either producing low-density channels or almost emptied holes with small clumps of matter remaining within. In both cases, we model such channels by a density-bounded component with low input ionization parameter (see Equation \ref{e34}).
Since the LyC emitters studied here are clearly dominated by very young stellar populations with ages $\la 3-4$ Myr, too young for significant feedback from supernovae \citep{Izotov_2018b}, we suspect radiation pressure and related feedback effects to be the main driver for the creation of these channels or holes, although the explanation may be more complex \citep[cf.][]{Hogarth_2020}.

{\em Density-bounded galaxy (DBG)}: Density-bounded galaxy with ionization-bounded clumps. The core of the galaxy is completely ionized by massive young stars, which produce a large number of ionizing photons exceeding what can be absorbed by the available nebular gas. This is represented by a high-$U$, density-bounded component, from which photons can directly escape at the edge. We add a low-ionization component to represent denser, ionization-bounded filaments or small clumps, which can naturally emit the \oi\ and \sii\ emission lacking in one-zone models (see section \ref{s_classical_mod}) and where the observed UV low-ionization lines are formed.
Both components have the same chemical composition and the same ionizing spectrum, emitted by a central source.

For a first approach,
we combined two arbitrary photoionization models with different ionization parameters: one with a high-ionization parameter $U_{\rm h}$ and one with a low-ionization parameter $U_{\rm l}$. The high- and low-ionization parameters were chosen "by hand" so that the nonleaking combined models accurately reproduce the upper part of the SDSS sequence on our diagnostic plots (see Sect.\ \ref{position_combined}) and the tight SDSS sequence in the O32-O13 and O32-S23 plots (Fig.\ \ref{fig_o_s} and \ref{fig_OS_combined}). We used $U_{\rm h} = 10^{-1}$ and $U_{\rm l} = 10^{-3.5}$. We briefly comment on this choice and an optimizing procedure below (Sect.\ \ref{ss_discuss1}).

We computed both scenarios for an identical stellar population of 1 Myr, a metallicity $\oh = 8$, and N/O=$0.1$, as above.
As our combined model is meant to represent a galaxy ionized by the same stellar population, we have to make sure that the photoionization models describing the two components also correspond to the same number of ionizing photons $Q_{\rm h}$ and $Q_{\rm l}$, respectively. This is simply achieved by scaling the predicted line luminosities of the $U_{\rm l}$ model with the factor $s=Q_{\rm h}$/$Q_{\rm l}$. The two components are then combined with relative weights, $\omega$ for the $U_{\rm h}$ component and (1-$\omega$) for the $U_{\rm l}$ component.
Any ratio of emission lines A and B of the combined model then becomes
\begin{equation}
    \frac{L^{A}_{\rm c}}{L^{B}_{\rm c}} = \frac{ \omega L^{A}_{\rm h} + (1-\omega) L^{A}_{\rm l}s}{\omega L^{B}_{\rm h} + (1-\omega) L^{B}_{\rm l}s}, \,\,
    {\rm where} \,\,
    s =Q_{\rm h}/Q_{\rm l}
    \label{e12}
.\end{equation}

For each scenario, we allowed the LyC escape fraction of the leaking component to vary from 0\% to 80\% in steps of 10\%, provided by the 3MdB\_17 database. The total \fesc\ of the two-zone model, is computed consistently from the relative weight of zones $\omega$, by defining
\begin{equation}
    f^{\rm c}_{\rm esc} = \omega f_{\rm esc}^{\rm h} + (1-\omega)f_{\rm esc}^{\rm l}.
    \label{e4}
\end{equation}{}
As only one component is leaking in each scenario, Eq.\ \ref{e4} simply becomes $f^{\rm c}_{\rm esc} = (1-\omega)f_{\rm esc}^{\rm l}$ in the DBC scenario and $f^{\rm c}_{\rm esc} = \omega f_{\rm esc}^{\rm h}$ in the DBG scenario.
In the absence of leakage ($\fesc=0$), the two scenarios are trivially identical.

\begin{figure*}[tb]{
   \centering
   \includegraphics[width=18cm]{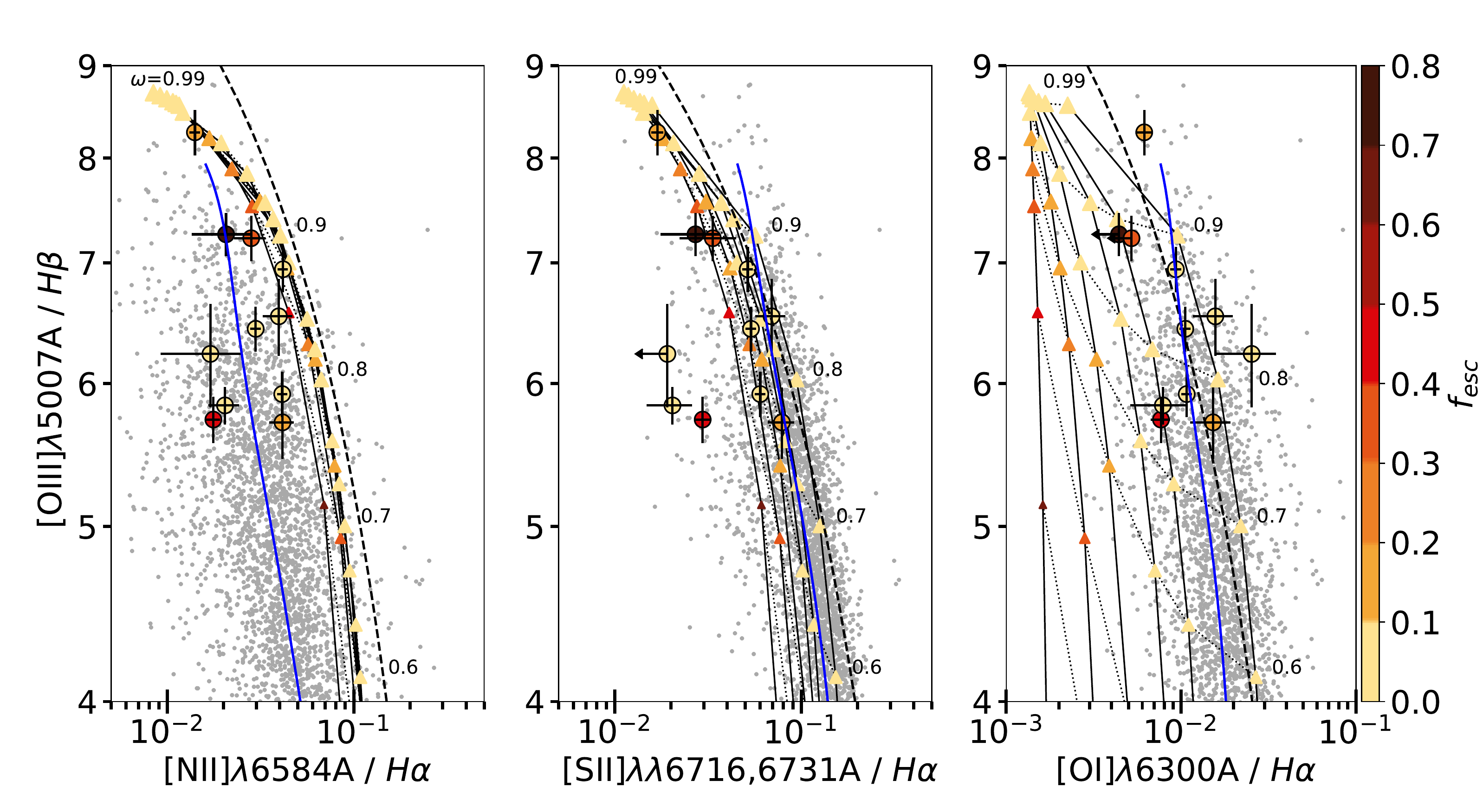}
   \includegraphics[width=18cm]{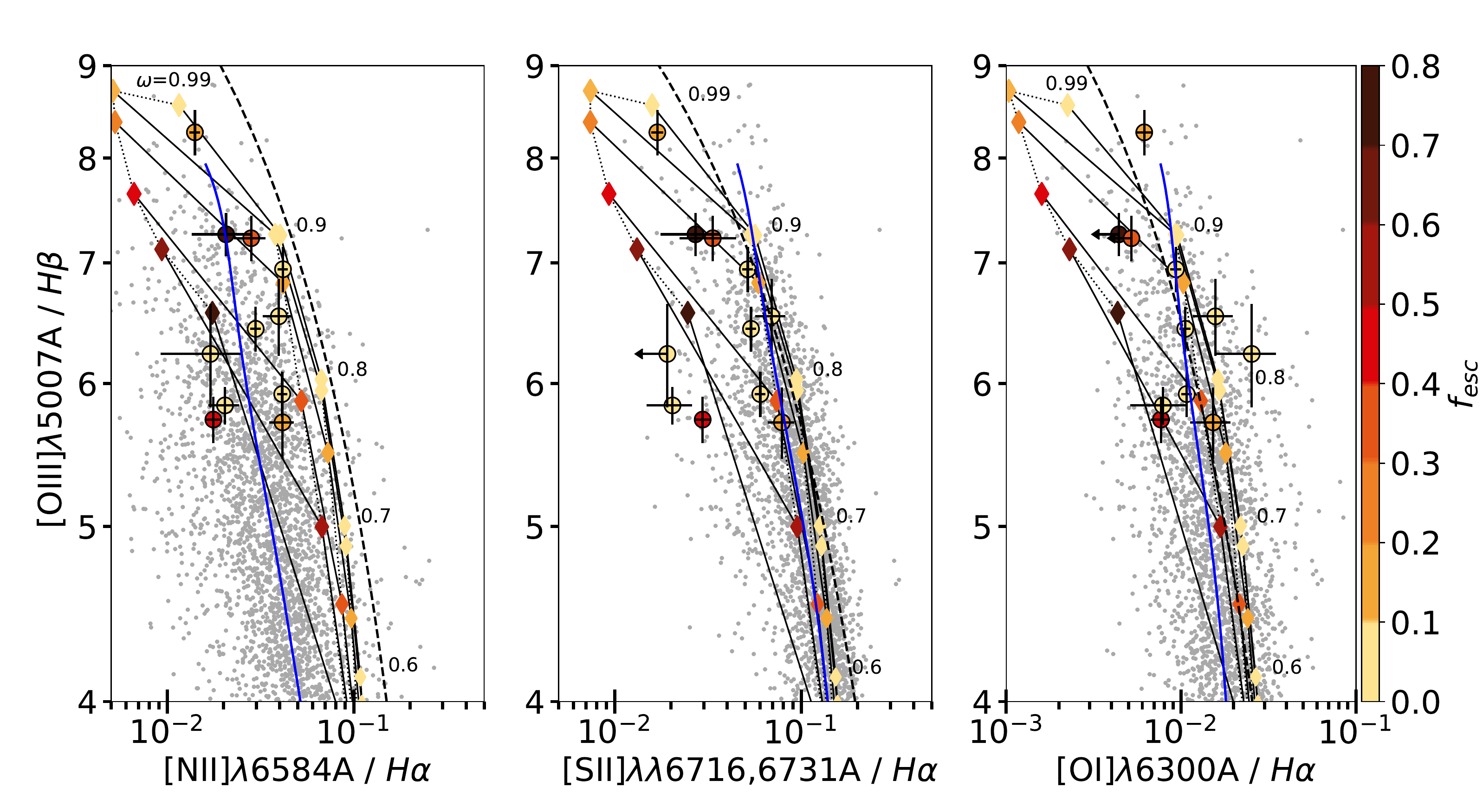}
   \caption{Classical diagrams (same as Fig.\ \ref{fig_BPT}) showing the predictions from the two-zone photoionization models. The upper and lower panels represent DBC and DBG models, respectively.}
   \label{fig_BPT_combined}}
\end{figure*}

\begin{figure*}[tb]{
   \centering
   \includegraphics[width=15cm]{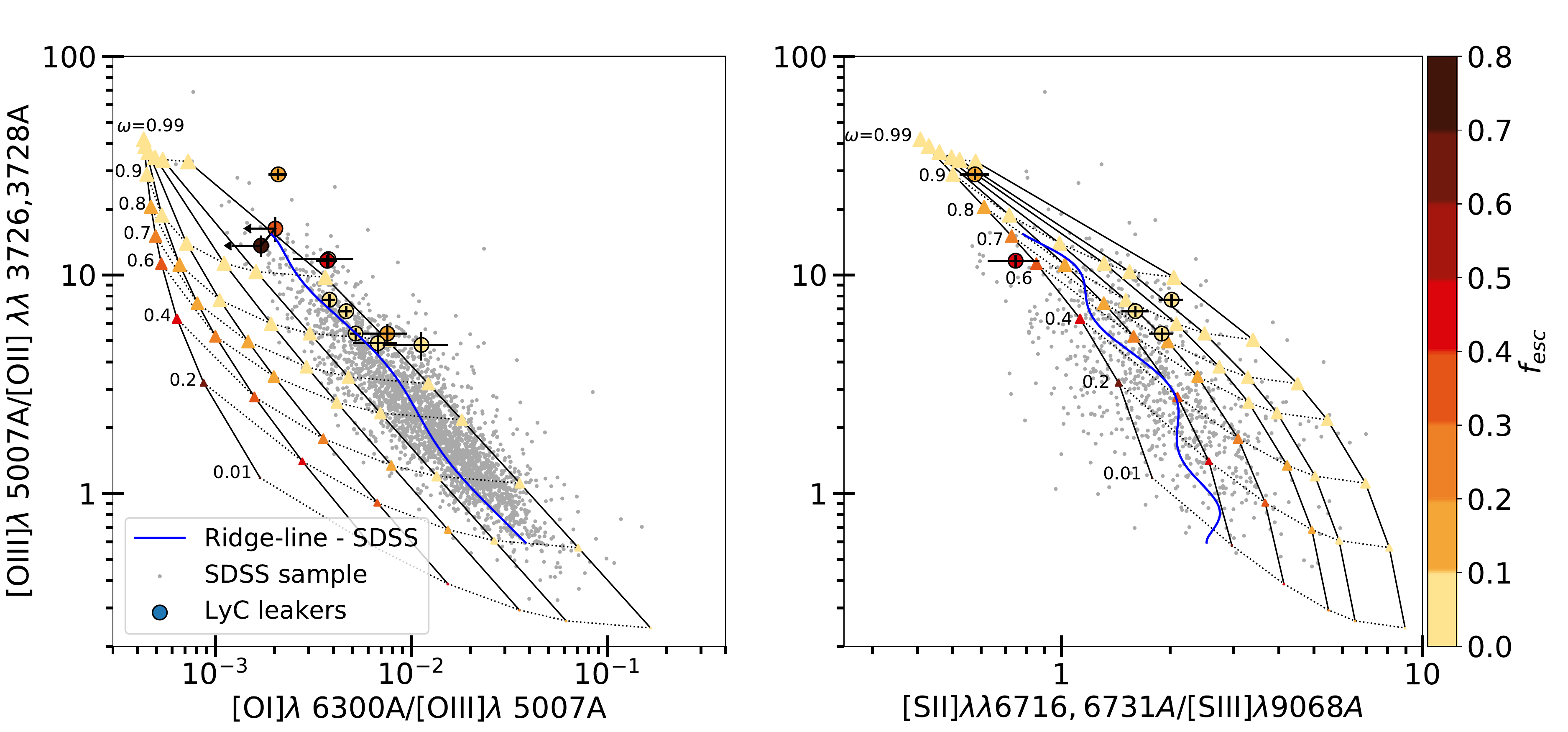}
   \includegraphics[width=15cm]{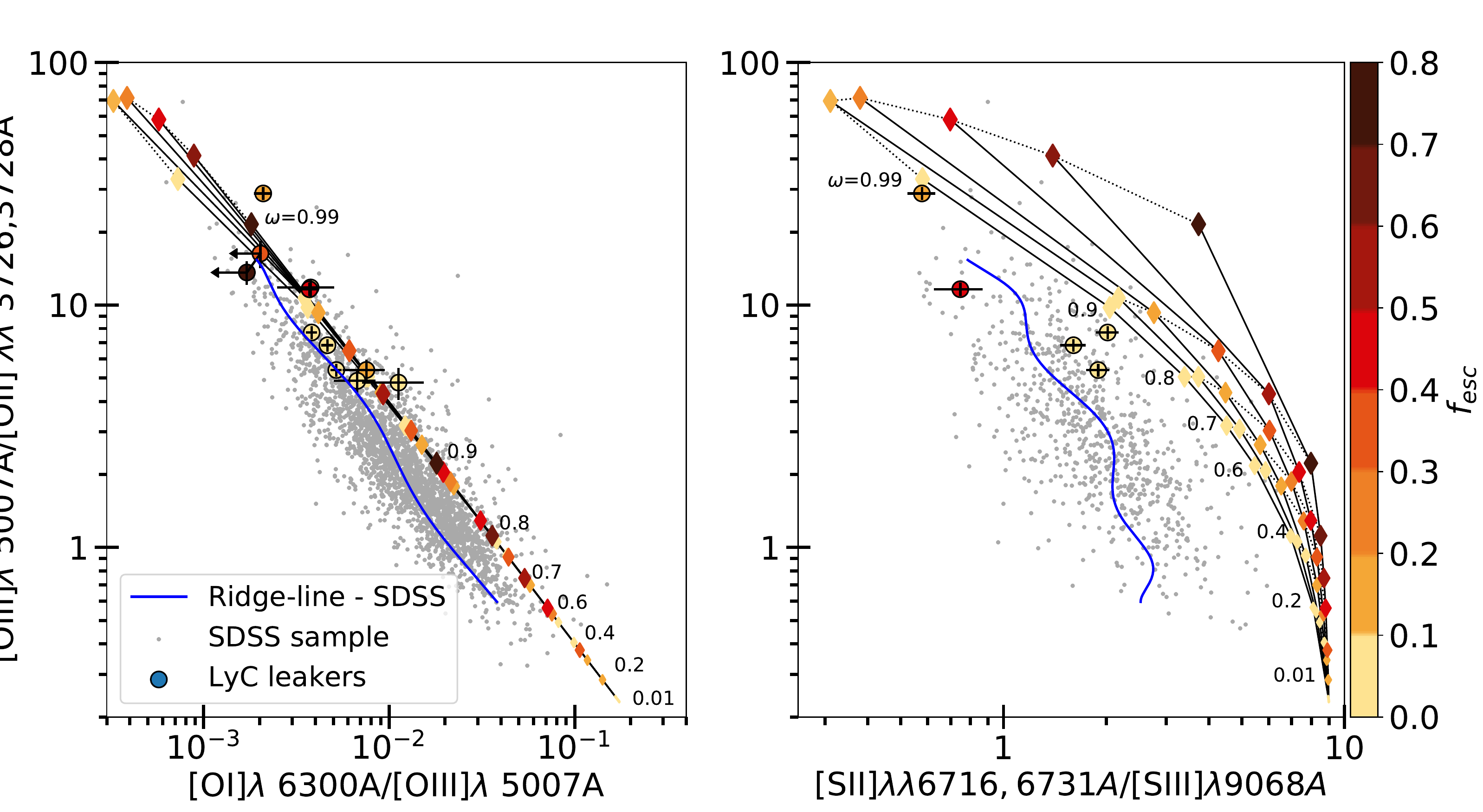}
   \caption{O32-O13 and O32-S23 diagrams (same as Fig.\ \ref{fig_o_s}) showing the predictions from the two-zone photoionization models. The upper and lower panels represent DBC and DBG models, respectively.}
   \label{fig_OS_combined}}
\end{figure*}

The resulting emission line ratios predicted from our two-zone models with varying relative weights and LyC escape fractions are shown in Figs.\ \ref{fig_BPT_combined} and \ref{fig_OS_combined}, allowing thus a comparison with the one-zone models (Figs.\ \ref{fig_BPT} and \ref{fig_o_s}). Nonleaking models are represented by the yellow symbols with 0\% escaping photons. In this case the two scenarios (DBC and DBG) are identical by construction, and their symbols (yellow triangles on the first row, and yellow diamonds on the second row) are thus superposed.
As mentioned before, we chose the ionization parameter of each component such that the combined nonleaking models adequately cover the upper part of the SDSS sequence.

\subsection{\oi\ and \Sii\ emission in normal galaxies explained}
\label{position_combined}
As we can see from Fig.\ \ref{fig_BPT_combined}, the nonleaking two-zone models follow a bended or sometimes straight ``mixing line'' from very high to moderate excitation (in \oiiil /\hb) and covering a wide range in the \nii/\ha, \sii/\ha, and \oi/\ha\ ratios.
Despite our somewhat arbitrary choice of models describing the individual zones, we already see that observed line ratios of the bulk of the SDSS galaxy sample (which primarily hosts galaxies with very low or zero \fesc) can be covered with the two-zone models  combining two different ionization parameters. In particular, there seems to be no difficulty in explaining the relatively high observed intensities of \Oi, and even the comparatively high values of \Sii/\Siii\ in normal galaxies, which cannot be reproduced by classical one-zone models (compare Fig.\ \ref{fig_o_s} and Fig.\ \ref{fig_OS_combined}). The inclusion of a second ISM component with a lower ionization parameter therefore
seems both essential and sufficient to solve the above-mentioned discrepancies of classical models when compared to normal galaxies.

\subsection{LyC emitters explained by two-zone models}
In contrast to the one-zone models, the combined two-zone models are able to cover the entire range of emission line ratios of the LyC emitters, as shown by Figs.\ \ref{fig_BPT_combined} and \ref{fig_OS_combined}. Most importantly, the models can---at least qualitatively---explain the observed \oi\ excess and the \sii\  deficiency (see Fig.\ \ref{fig_s2} and Fig.\ \ref{fig_o321}), which characterizes the leakers, and which cannot be understood by one-zone models (see Sects.\ \ref{obs_trends} and \ref{spectral_diag}).
Furthermore, the two-zone models account for the escape of LyC photons, and are thus consistent with this important observational property. Below we further explore the consistency of different line ratios and the extent to which the models also correctly reproduce the observed LyC escape fractions.

Let us now examine how the predictions from the two-zone models in classical diagrams vary with LyC escape fraction. In the O3H$\beta$-N2H$\alpha$ diagram (Fig.\ \ref{fig_BPT_combined} left), most of the models are superposed quite independently of their \fesc\ value. 
This is due to the fact that \nii\ originates from the highly excited region in the inner part of the H~{\sc ii} region, which remains unaffected when trimming the outskirts of the galaxy. Only the most density-bounded models from the DBG scenario, where the highly ionized component dominates almost all the galaxy ($\omega =99\%$) and is leaking, affect the \nii/\ha\ ratio when \fesc\ exceeds $10\%$. 
In contrast, the \sii\ and \oi\ line ratios (shown in the middle and right panels) are very sensitive to LyC leakage. We note that these emissions primarily arise from the low-ionization component, that is,\ the one leaking in the DBC scenario (upper panel). This explains why these line ratios are barely affected in the DBG scenario (lower panel), even for high escape fractions. On the contrary, leakage in the low-ionization component (DBC scenario) rapidly decreases \sii\ and \oi: only low escape fractions ($\fesc \la 10$\%) are compatible with the observed \oi\ emission of the leakers in this scenario.

The same trend is visible on the O32-O13 plot (Fig.\ \ref{fig_OS_combined}). The SDSS sequence seems bounded by $\fesc\sim0-10$\% for the DBC scenario (first row). Such values are sufficient to explain the observed escape fraction of some of the leakers, represented by yellow circles, but it fails to reproduce higher leakage of ionizing photons, represented by the orange to brown circles. 
Meanwhile, the DBG scenario can produce escape fractions up to $80\%$ without  strongly decreasing the emission of \oi\ and \sii.
On the other hand, the O32-S23 plan seems to be better covered by models from the DBC scenario, while the DBG scenario tends to overestimate the \sii/\siii\ ratio. For the five leaking galaxies with \siii\ measurements, the observed escape fractions seem consistent with predictions from the DBC scenario, even for the most leaky galaxy J1154+2443 (red circle, $\fesc=46$\%). Indeed, we note that this galaxy stands on the left-hand side of the median line, as expected for the highest escape fractions produced by the DBC scenario; it falls between two model points with escape fractions of 30--40\%, which is roughly consistent with the observed value.
We also note that models with a weight $\omega \la 0.7$ fail to reproduce the leakers studied here. This is easily understood if we keep in mind that these galaxies show high excitation, as indicated by their high O32 ratio: decreasing the contribution $\omega$ of the $U_{\rm high}$ region below a certain threshold produces two-zone models that are dominated by the low ionization ($U_{\rm low}$) component, and hence are unable to reproduce the observed line ratios of the leakers.

% % % % % % % % % % % % % % % % % % % % % % %
\subsection{Links with \fesc} 
\label{s_links_fesc}
\begin{figure*}[tb]{
   \centering
   \includegraphics[width=15cm]{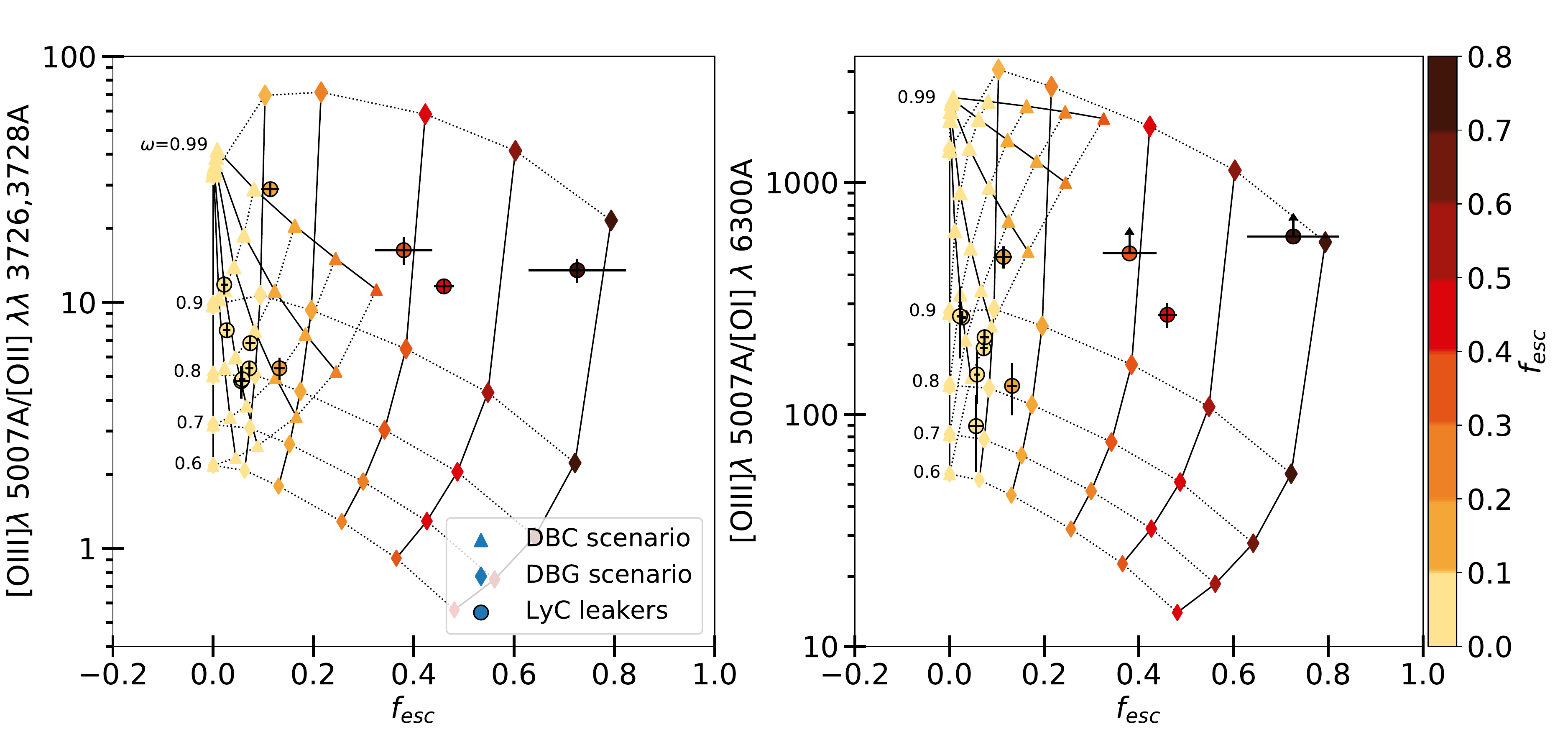}
   \caption{Same as Fig.\ \ref{fig_o32_fesc_BOND} showing the predictions for the two-zone photoionization models.}
   \label{fig_O321_fesc_combined}}
\end{figure*}

In Fig.\ \ref{fig_O321_fesc_combined} we now revisit the predicted O32 and O31 line ratios in models with LyC escape presented earlier for one-zone models (see Fig.\ \ref{fig_o32_fesc_BOND}). As already mentioned and clearly shown in this figure, the two-zone models solve the \oi\ problem of simple models, and they cover the entire range of observed line ratios of the LyC emitters. A comparison of the two panels also shows that, at least approximately and without adjustment of the components of the two-zone models, the models are capable of consistently reproducing the two oxygen line ratios and the observed escape fraction.
Figure 10 also shows that LyC emitters with very high $\fesc \ga 20$\% can only be reproduced by our DBG scenario (where the high $U$ component is leaking), whereas both scenarios are possible for the galaxies with lower escape fractions. 

\begin{figure*}[tb]{
   \centering
   \includegraphics[width=15cm]{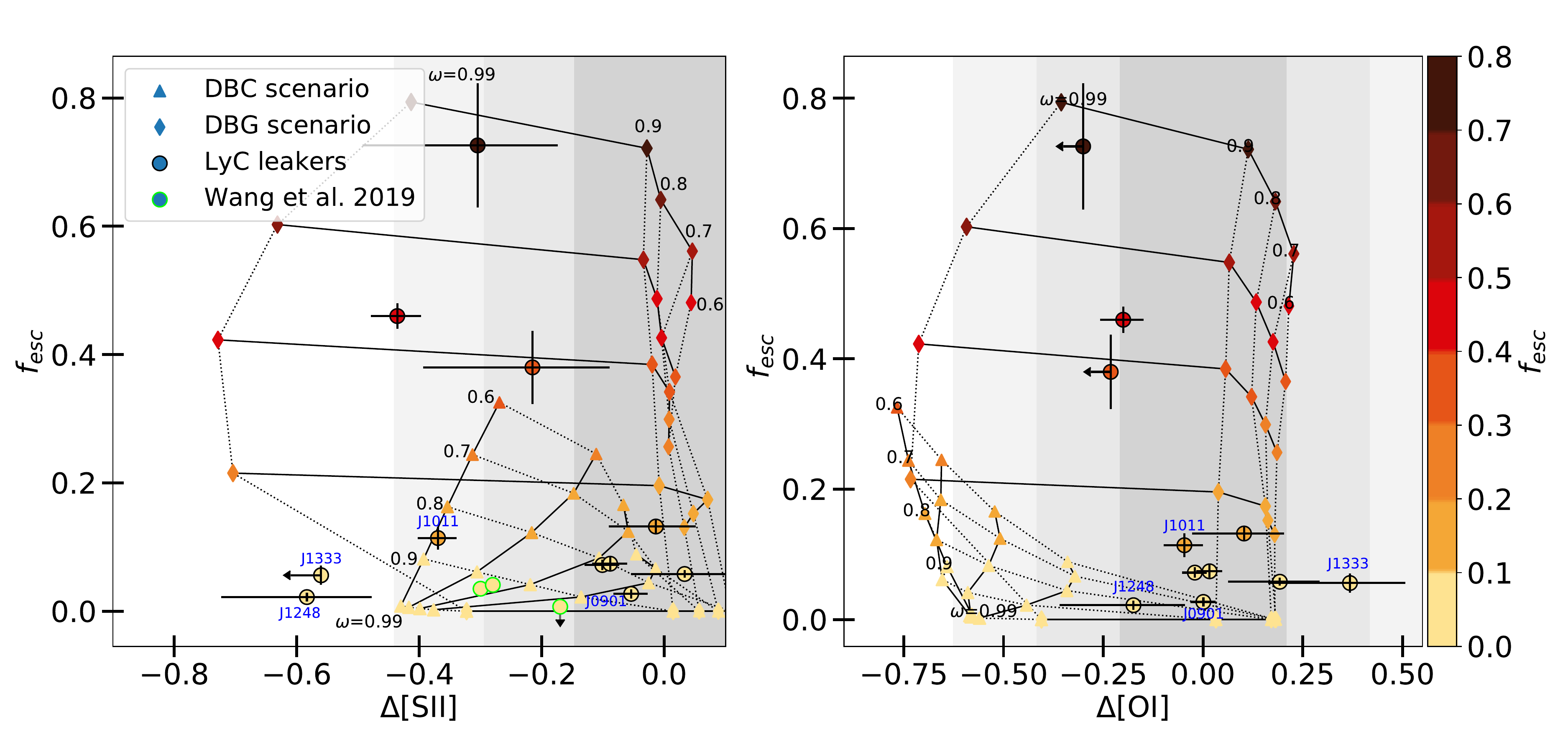}
   \caption{LyC escape fraction as a function of the deficiency $\Delta$\sii\ in \sii\ (left) and $\Delta$\oi\ (right). The gray-shaded areas represent dispersion of the SDSS comparison sample at 1, 2, and 3~$\sigma$. The three strongest leakers ((J1256+4509, J1154+2443 and J1243+4646, in increasing order of $f_{esc}$) are represented by orange to brown circles in the upper part of the plots. The other galaxies that are discussed in Section \ref{s_links_fesc} are labeled in blue.}
   \label{fig_deltas}}
\end{figure*}

Figure \ref{fig_deltas} (left) illustrates the evolution of \fesc\ as a function of $\Delta$\sii. 
This plot suggests that a sufficiently strong \sii\  deficiency (e.g.,\ $\Delta$\sii\ $\la -0.2$) effectively picks out LyC leakers,
although not all of them. Indeed, several LyC emitters are also located within the normal range of the comparison sample 
(presumably dominated by nonleakers), as also pointed out by \cite{Wang_2019}. 
However, using  $\Delta$\sii\ to determine the escape fraction is not obvious. Compared to the dispersion of our comparison sample represented by the gray-shaded areas, 6 out of 11 leakers show a deficit in \sii\ larger than 1$\sigma$ (5 out of 11 with deficit larger than 2$\sigma$). If we include the 2 leaking galaxies from \cite{Wang_2019}, this amounts to 8 out of 13 galaxies with a clear \sii\  deficit at 1$\sigma$ (6 out of 13 at 2$\sigma$).
The three strongest leakers (J1256+4509, J1154+2443 and J1243+4646), which have $\fesc \ga 30$\% all exhibit a large deficiency in \sii\ of at least 0.2 dex from the ridge line (> 1$\sigma$). J1011+1947, with an intermediate escape fraction of 11.4\% also shows a large \sii\ deficit. However, the most \sii-deficient galaxy in our sample (J1248+4259) and J1333+6246, which is undetected in \sii, have escape fractions of only a few percent. This could mean that the observed \fesc\ for these galaxies is underestimated, for example due to the fact that LyC radiation escapes only along other, unobserved lines of sight.
Interestingly, in the analysis of the UV absorption lines of more than 30 low-$z$ galaxies from \cite{Gazagnes_2020}, J1248+4259 stands out as an outlier with an exceptionally strong \lya\ emission
(high equivalent width) despite showing a large covering fraction in the H~{\sc i} absorption lines. Demonstrating whether or not any of these sources indeed emit more ionizing photons in other directions seems difficult.
In any case, our analysis confirms that the method proposed by \cite{Wang_2019} could be a promising way to identify new leaking galaxies, although the determination of \fesc\ from the \sii\  deficit remains challenging. 

From Fig.\ \ref{fig_deltas} (left) we notice once more that only the DBG scenario is able to reproduce the more extreme leakage above 20\%. The DBC scenario is able to cover the space where most of the galaxies with small $f_{esc}$  reside, including the \sii-deficient galaxies selected by \cite{Wang_2019} (green color along the contours of circles). However, we note that the two galaxies with the strongest \sii\ emission ($\Delta$\sii\ close to zero) stand outside of the contours of the DBC scenario. This indicates that leakage powered exclusively by channels of low ionization is unable to simultaneously produce strong \sii\ emission and escaping LyC photons. This comes from the fact that in this scenario leakage is allowed exclusively by trimming the low ionization component, which always results in decreasing \sii\ emission. This could either suggest that density-bounded LyC could still be the preferred mechanism even for rather small (\la 10\%) escape fractions, or that a more complex geometry needs to be considered to allow leakage without systematically decreasing \sii. 

In Fig.\ \ref{fig_deltas} (right) we show $\Delta$\oi\  calculated from the O3H$\beta$-O1H$\alpha$ plots and defined earlier (see Fig.\ \ref{fig_BPT}, Sect.\ \ref{obs_trends}). Positive $\Delta$\oi\ corresponds to galaxies showing an excess in \oi/\ha\ compared to the SDSS median line, while a negative $\Delta$\oi\ marks an \oi\  deficiency. Interestingly, we notice a visible trend of \fesc\ relative to $\Delta$\oi\ in the observations: while the most \oi-deficient galaxies and the two undetected galaxies ($|\Delta$\oi|> 1$\sigma$) exhibit the largest escape fractions, the observed \fesc\ decreases for increasing $\Delta$\oi.
This could indicate that \oi\ emission, which traces the photo-dissociated interface between ionized and atomic gas or cold dense clumps in the H~{\sc ii} region might be linked to the disappearance of the remaining Lyman continuum, leading to lower escape fractions. However, we notice  that two galaxies (J1248+4259 and J0901+2119) that are \oi-deficient exhibit very small escape fractions ($\fesc \sim 2\%$). This indicates that other effects are likely to cause the drop of Lyman continuum even in \oi- deficient environment. However, the overall trend shown by the observations in Fig.\ \ref{fig_deltas} suggests that selecting \oi-deficient galaxies could also be an effective way to target galaxies with high leakage and might correlate better with \fesc\ than \sii\  deficiency. On the other hand, the \sii\ lines are generally stronger than \oi, making the former easier to measure.

In Fig.\ \ref{fig_deltas} (right) the contours from the DBC scenario include only the two galaxies with the lowest escape fractions. For the same reasons as for \sii\ emission, leakage through low-ionization channels tends to be associated with low \oi\ emission, that is, $\Delta$\oi\ \la 0. The galaxies that are not \oi\  deficient are not well reproduced by the DBC scenario, especially the galaxy standing on the extreme right-hand side whose \oi\ emission exceeds that of all the models considered here. This might suggest that the physics of \oi\ emission is not well taken into account in our two-component models, and might require a boost from an additional component (see discussion in Section.\ \ref{s_discuss}). However, we note that apart from the systematic offset toward lower \oi\ emission, the DBC scenario qualitatively reproduces the observed trend of increasing \fesc\ with decreasing $\Delta$\oi. The DBG scenario accurately reproduces all the $\Delta$\oi\ and \fesc\ of the leakers (except for the strongest \oi\ emitter) but $\Delta$\oi\ is independent from the escape fraction in this scenario; the observed variations in $\Delta$\oi\ only come from the decrease of \oiii\, which vertically moves the models closer to the ridge line but without affecting \oi\ emission. 

We conclude that $\Delta$\sii\ and $\Delta$\oi\ could be interesting proxies to estimate the escape fraction but only in the case of ionization-bounded leakage (DBC scenario). In the case of density-bounded leakage (DBG scenario), likely producing the highest escape fractions, \oi\ and \sii\ seem uncorrelated with \fesc, and other diagnostics sensitive to density boundaries are to be preferred (see, e.g., Fig.\ \ref{fig_o_s} and \ref{fig_OS_combined}). 
It seems that \oiii/\oii, $\Delta$\sii,\ and \oi\  excess are good hints for LyC leakage, but there still seems to be other independent dimensions, which probably incorporate several physical parameters such as the nature of the ionizing cluster, its evolutionary stage, the gas content of each galaxy, its star formation rate, and metallicity, all of which could produce other combinations for a LyC leaking cluster \citep[e.g.,][]{Wang_2019}.

%%%%%%%%%%%%%%%%%%%%%%%%%%%%%%%%%%%%%%%%%%%%%%%%%%%%%%%%%%%%%%%%%%%%%%
\section{Discussion}
\label{s_discuss}

We now briefly discuss further improvements, potential caveats, implications, and future prospects resulting from the present work.
% % % % % % % % % % % % % % % % % % % % % % %
\subsection{Can the LyC escape fraction be detected and measured from the optical emission lines ?} \label{ss_discuss1}

In this study, we used the observed LyC escape fractions and optical emission lines to build simple but consistent two-zone
photoionization models of the LyC emitters, and to discriminate between different leakage scenarios. An obvious further step would be to use these two-component models to infer the expected escape fraction from the emission line observations only. Such indirect methods are badly needed in order to optimally exploit upcoming JWST observations of rest-frame optical emission lines for galaxies during the epoch of reionization.

A first automated routine was implemented in an effort to extract best-fit values for the parameters of the two-component models. Although very promising, this method requires a robust implementation and a careful choice of tracers, which we postpone to future work. We present here the first trends obtained from this study and discuss possible caveats of this method.

We ran a $\chi^2$-minimization routine on a subgrid of our combined models, which includes the ones presented in Sect.\ \ref{s_composite_models}. We defined a four-dimensional grid with the following free parameters: \uhigh\ and \ulow (varied from $10^{-4}$ to $10^{-1}$), the weight of the component of high ionization $\omega$, and the escape fraction of the leaking component $f_{\rm esc}^{\rm holes}$ (both varied between 0 and 1).  
As observational constraints we used six line ratios: the four ratios used in the classical diagrams (Fig.\ \ref{fig_BPT_combined}), and two ratios from the O32-O13 diagrams (Fig.\ \ref{fig_OS_combined}). We did not include upper limits. As our prescription for $f_{\rm esc}^{\rm combined}$ is fully determined by the parameters of each model (see Eq.\ \ref{e4}), we could also determine the corresponding value of the total LyC escape fraction for the $\chi^{2}$ minimum. We also examined the weight of the ionization-bounded component $\omega$ in the DBC scenario (respectively $(1-\omega)$ in the DBG scenario) to compare it to the UV covering fractions deduced from the Lyman series absorption lines \citep{Gazagnes_2018, Chisholm_2018,Schaerer_2018}. Hereafter, we summarize some interesting trends that merit further examination:

\begin{itemize}
\item The models minimizing $\chi^{2}$ always favor two distinct values for the ionization parameters \uhigh\ and \ulow. Inspection of their marginalized probability distribution functions shows that typically $\uhigh \sim 10^{-1}-10^{-2}$, and $\ulow \sim 10^{-3}-10^{-4}$. We interpret this as confirmation of our conclusion in Sect.\ \ref{s_composite_models} that two-component models do a better job at reproducing the observed line ratios than single-component models.

\item Regardless of the considered scenario, the weight of the high-ionization component is always much higher than that of the low-ionization one ($\omega \sim 0.8$). This means that no matter which component is leaking, the high-ionization component always dominates the total emission, in the same proportion of about $~80\%$ (respectively $~20\%$ for the low-ionization component). This suggests that the evolution from one scenario to the other could result from a continuous evolution \citep[cf][]{Kakiichi_2019} leaving the global proportions of the structure of the galaxy unchanged. However, we note that the two scenarios produce very different covering fractions. 
We note that for galaxies with a low escape fraction (\fesc$\la~10\%$), the $\omega$ obtained from the best-fitting model has a value comparable to the covering fractions, $C_f$, obtained from  analysis of absorption lines \citep{Gazagnes_2018, Chisholm_2018,Gazagnes_2020}. This is consistent with what we expect from the DBC scenario, where the covering fraction should be close to $\omega$ but cannot be reconciled with the DBG scenario, where $C_f \sim (1-\omega),$  which is a better fit for some of these galaxies. 
On the other hand, for strong leakers ($\fesc \ga 10$\%), we find $C_f < \omega$. This strengthens the conclusion in Sect.\ \ref{s_links_fesc} that the DBC scenario is unable to reproduce strong leakage, as it predicts too high a covering fraction, and therefore overly low escape fractions. However, we note that the DBC scenario sometimes better reproduces the six line ratios we considered, even for strong leakers.

\item Finally, the escape fraction obtained for the best-fit models allows us to distinguish the four galaxies with the strongest leakage in our sample. The best-fit models predict $\fesc \approx 0$
for all the weak leakers but one, and nonzero escape fractions ($\fesc \sim 0.6-38\%$) for the galaxies with high escape fractions. We note that the highest predicted escape fraction corresponds to J1243+4646, the strongest leaker in our sample, although the predicted value of \fesc\ is a factor of approximately two below the observed value.
\end{itemize}

Briefly, this exercise supports our conclusion in Sect.\ \ref{s_links_fesc} that the geometry of weak leakers resembles that of the DBC scenario, and strong leakers more the DBG scenario. However, we find that robustly distinguishing between the two scenarios based only on emission lines is not easy. Furthermore, accurate predictions of \fesc\ from the optical emission line ratios do not appear to be straightforward.
This will probably require a careful choice of the relevant line fluxes or ratios to be used in such an analysis, a proper treatment of upper limits and uncertainties, a robust statistical method to overcome the caveats of a simple $\chi^{2}$ minimization, 
the exploration of SEDs allowing also for different ages, and other improvements, which are beyond the scope of the present publication.

\subsection{Low ionization line diagnostics: progress and caveats} \label{ss_discuss2}

In Sect.\ \ref{s_classical_mod}, we discussed the inadequacy of simple models to  simultaneously reproduce the high and low ionization lines, as revealed for example in the O32-O13 and O32-S23 diagnostics (Fig.\ \ref{fig_o_s}). Specifically, we emphasized the offset of BOND models in O32-S23, which systematically underestimate the \sii/\siii\ ratio. This offset of standard photoionization models has been shown in previous work \citep[e.g.,][]{Perez-Montero_2009, Kehrig_2006}, in which such models fail to explain the hardening in ionizing radiation shown by most low-mass, low-metallicity H~{\sc ii} galaxies in the O32-S23 plane.
Problems with the sulfur line strengths have often been reported in the past
\citep[see, e.g.,][and references therein]{Garnett_1989,Kewley_2019,Mingozzi_2020}. The shortcomings of 
several Cloudy and Mappings models to reproduce the \sii/\siii\ ratio have for example recently been shown by \cite{Mingozzi_2020}.
Considering that this only seems to affect sulfur lines, these latter authors suggest that this is due to limitations in stellar atmosphere modeling and/or the atomic data for sulfur. Based on the observation that the discrepancy holds for single H~{\sc ii} regions \citep[from][]{Kreckel_2019}, \cite{Mingozzi_2020} exclude the role of diffuse ionized gas as a cause of this issue. However, in our study, we see in Fig.\ \ref{fig_OS_combined} that adding a low-ionization component, which provides additional emission of \sii\ without producing \siii, helps to solve this discrepancy. 
We conclude that for both our reference sample of star-forming galaxies and the LyC-emitting galaxies, the addition of such a low-ionization component is sufficient to reproduce the observed \sii/\siii\ ratios. Furthermore, this component also contributes to low ionization emission of oxygen, simultaneously reproducing the observed line strength of \Oi\ and lines tracing the other ionization stages of oxygen. However, the exact nature, spatial distribution, and other properties of such a component remain to be understood.

While this success of our two-zone models is notable and certainly calls for further studies and applications,
it must be noted that other physical processes can also play a role for the low-ionization lines. For example, it is well known that shocks in the ISM or the presence of sources with hard ionizing spectra (far-UV and X-ray emission)
lead to enhanced emission of highly ionized species (e.g.,\ [Ne~{\sc v}], [Fe~{\sc v}], [O~{\sc iv}], \heii, and others) and also of low-ionization lines
\citep[e.g.,][]{Schmitt_1998,Stasinska_2015, Plat_2019}. 
The effect of a contribution from shock models with different magnetic field strengths (tuned to reproduce a typical \heii/\hb\ ratio) on the O32-O13 diagram is shown in \cite{Stasinska_2015}. Clearly, some model combinations of shock with both density- and ionization-bounded H~{\sc ii} regions can be found that reproduce the three ionization stages of oxygen. How these models would affect the sulfur lines, and whether or not they could also be consistently reproduced remains to be seen. Given different degeneracies and uncertainties it seems currently difficult to exclude a contribution from shocks \citep[see][]{Stasinska_2015}.
Similarly, a weak AGN component (contribution $\sim 3$\% of the hydrogen ionizing photons) cannot be excluded based on the observed emission line
diagrams for the galaxies of our reference sample \citep{Stasinska_2015}. Again, the sulfur line ratios used here may provide an additional handle on these questions. 

Finally, the low ionization lines are also affected by diffuse ionized gas (DIG) in galaxies, as recently discussed by
\cite{Mingozzi_2020}, \cite{Shapley_2019}, \cite{Sanders_2020}, and others. 
In the LyC emitters studied here, the contribution of the DIG would presumably be low, if we adopt the observed scaling of the DIG contribution with 
surface density of star-formation \cite[e.g.,][]{Oey_2007,Shapley_2019}, since the LyC emitters are compact by selection, and have a very high SFR surface density \citep{Izotov_2018b}.

Regardless of the exact origin of emission of the low-ionization lines (such as \Oi\ and \Sii) in LyC emitting galaxies,
the UV observations of these galaxies clearly reveal the presence of both low-H~{\sc i}-column-density (\nh) lines of sight towards the observer
allowing the escape of LyC radiation and lines of sight with high \nh, where other neutral and low-ionization metal absorption lines are also formed
\citep{Gazagnes_2018,Chisholm_2018,Gazagnes_2020}. Therefore, their ISM contains at least two ``zones'' or\ regions
with different conditions, even if we do not know the exact spatial scales on which this happens. Furthermore, it is very plausible that properties that influence the ionization parameter (e.g.,\ density, clumpiness, filling factor) differ between these regions.
This justifies our two-zone approach using two different ionization parameters, \uhigh\ and \ulow, for the LyC leakers.
Whether or not and to what extent such an approach is also motivated for ``normal'' galaxies (most of which are presumably not LyC emitters) remains to be seen.
In any case, given the ubiquitous complexity of the ISM in resolved star-forming galaxies it is probably not surprising that simple 1D photoionization models
cannot reproduce all the observables. 

\subsection{Future improvements}
Some of the methodology used and some of the questions addressed here certainly call for other applications and for future improvements.
Indeed, although photoionization models are frequently used to predict and analyze emission lines in star-forming galaxies, they are generally 1D spherical or plane parallel models computed with Cloudy or similar codes, which then allow for exploration in a wide range of parameter space in terms of ionization parameters, radiation field, abundances and so on.
Such models, which typically assume a single source of ionizing radiation and a very simple geometry, in contrast to observations of integrated galaxy spectra, are obvious oversimplifications of the true complexity and diversity of the radiation field, ISM, and geometry.  

Relatively few more highly complex models have so far been used for comparisons with observations.
\cite{Stasinska_2015} for example explored the combination of stellar photoionization and other emission sources (AGN, shocks, and others) with variable but arbitrary contributions, and confronted them with a large sample of star-forming galaxies from the SDSS.
\cite{Cormier_2012,Cormier_2019}  used one- and two-component models to consistently analyze emission lines originating in the ionized and neutral ISM, as well as in adjacent PDRs. These have been compared to detailed multi-wavelength observations of 39 nearby dwarf galaxies. 
A very complex tailored model including up to five components, allowing also for ionization- and density-bounded parts, has been constructed for the well-known metal-poor galaxy I Zw 18 by \cite{Lebouteiller_2017}. Clearly other models can and need to be envisaged, although sufficient independent observational constraints are necessary to avoid degeneracies in a multi-dimensional parameter space. 

Another approach is to use high-resolution hydrodynamic simulations of galaxies, ideally including radiation transfer and a proper treatment of the ISM, to predict the resulting emission lines and compare them to observations. Such models, with fairly different degrees of sophistication have for example been presented by \cite{Shimizu_2016}, \cite{Hirschmann_2017}, \cite{Katz_2019}, \cite{Wilkins_2020},
and by \cite{Pellegrini_2020} for smaller scales.
These simulations are able to capture several observational trends and allow useful predictions, in particular for high-redshift galaxies. However, direct and detailed comparisons to 
observations are not trivial to undertake or to interpret. For example, \cite{Katz_2020} 
predicts optical and IR emission lines of LyC leakers and nonleakers, which show good agreement for the \Oiii, \Oii, and \hb\ emission lines when compared to observations. In contrast, their predictions for \Sii/\ha\ and \oiii/\hb\ are quite different from those observed in low-$z$ LyC emitters and in $z \sim 2-3$ galaxies. Understanding the origin and importance of such a discrepancy is nevertheless not straightforward. Different and complementary modeling approaches should help to develop more robust diagnostics and should therefore provide better insight into the properties of the ISM, radiation field, geometry, and other properties of star-forming galaxies and LyC emitters.

In parallel, new observations, such as those from the HST Low-$z$ LyC survey (PI A.\ Jaskot), will provide for the first time a statistical sample of more than 60 galaxies at  $z \sim 0.2-0.4$  with direct LyC observations. This sample of confirmed LyC emitters and also nonLyC emitters should represent a useful base to test several of the trends observed here, and an ideal sample for detailed comparisons with emission line models from different approaches. The first comparison with simple two-zone models and a small sample presented here is very encouraging, but incomplete. Nevertheless, there is hope that optical emission lines can to some extent be used to identify LyC emitters, and will ideally allow determination (even approximate) of the escape fraction of ionizing photons
from these galaxies. Clearly, future improvements on these lines will be very welcome and of great interest, especially in view of the upcoming launch of the JWST, which will routinely observe rest-frame optical emission lines out to the highest redshifts.

%%%%%%%%%%%%%%%%%%%%%%%%%%%%%%%%%%%%%%%%%%%%%%%%%%%%%%%%%%%%%%%%%%%%%%
\section{Conclusion}
\label{s_conclude}
We analyzed the main optical emission lines of low-$z$ LyC-emitting galaxies (leakers) discovered recently with the HST \citep{Izotov_2016a,Izotov_2016b, Izotov_2018a,Izotov_2018b,Wang_2019} in the classical excitation diagrams involving the \oiii/\hb,
\nii/\ha, \sii/\ha, and \oi/\ha\ line ratios, as well as other diagnostic diagrams involving three ionization stages of oxygen (using \oiii, \oii, and \oi) and two of sulfur (using \sii\ and \siii\ observations).
In comparison with the SDSS sample of star-forming galaxies with similar metallicities (\oh $\sim 7.6-8.2)$, we found several peculiarities of the LyC emitters, including an excess of the \oi/\oiii\ ratio for a fraction of the leakers \citep[as already pointed out by][]{Plat_2019}, and an \oi\  deficiency for those with the highest LyC escape fractions ($\fesc \ga 20\%$). We also confirm a deficit in \sii\  found recently by \citet{Wang_2019}, and suggest that combined measurements of the \oi\ and \sii\ lines could provide a good indicator for strong LyC emission.

We then compared the observations to simple one-zone and two-zone photoionization models accounting for the escape of ionizing radiation in order to construct, for the first time, consistent models which reconcile their observed emission lines and the measured LyC escape fractions. The main results from this analysis can be summarized as follows:
\begin{itemize}
    \item Classical one-zone photoionization models underpredict the low-excitation line emission of \oi\ and \sii\ arising from the outskirts of H~{\sc ii} regions; this underprediction is strong for density-bounded models with LyC escape, and also exists for ionization-bounded models (see Sect.\ \ref{s_classical_mod}), although it is less strong.
    \item Two-zone models, combining regions with a high- and low-ionization parameter (\uhigh, \ulow), where one of which is density-bounded, allowed us to resolve these discrepancies and to reproduce the observed line ratios of the LyC emitters. Furthermore, the models show LyC escape fractions comparable to those measured directly from the UV LyC observations. 
    \item From the two-zone models we also found a possible dichotomy between galaxies with different LyC escape fractions. Indeed, the models indicate that galaxies with $\fesc < 10$\% can be explained by the presence of low-column-density, low-filling-factor regions or channels with a low-ionization parameter through which the ionizing photons escape. The observed \oi\ emission can originate from these filaments/channels, in what we called the DBC scenario (Fig.~\ref{fig_scenario}). 
    For much higher \fesc, our photoionization models show that LyC leakage must occur from the region with high-ionization parameter (DBG scenario) in order to explain the observed emission line ratios. 
    %\item 
\end{itemize}

Our two-zone models are inspired by and compatible with the ``picket-fence'-type model, which successfully reproduces the observed UV absorption lines of hydrogen and low-ionization metal lines of the LyC emitters, as demonstrated by the studies of \cite{Gazagnes_2018, Gazagnes_2020}. Our finding of two different scenarios, favored by LyC emitters with a low or high escape fraction, is also compatible with the recent analysis of \cite{Gazagnes_2020} combining information from the \lya\ line profiles and absorption lines. Our results are also consistent with the numerical simulations of \cite{Kakiichi_2019}, which suggest that the mechanism powering LyC escape may evolve in a continuous way from the DBC scenario, producing low escape fractions, to the DBG scenario, which appears as a necessary configuration to reach high levels of leakage. 

Finally, we have discussed the possibility of improved analysis tools to predict escape fractions based on optical emission lines and using two-zone photoionization models (Sect.\ \ref{s_discuss}). 
We conclude that robust predictions would require multi-phase and multi-sector modeling, including some density-bounded components to successfully capture the complexity of the ISM structure that leads to photon leakage. We also suggest that low-excitation tracers from the outer part of H~{\sc ii} regions such as \oi\ and \sii\ lines should be included in the list of relevant lines to study the morphology and dynamical state of leaking and nonleaking galaxies.

%%%%%%%%%%%%%%%%%%%%%%%%%%%%%%%%%%%%%%%%%%%%%%%%%%%%%%%%%%%%%%%%%%%%%%%%%%%%%%%%%
\begin{acknowledgements}
We acknowledge the use of photoionization models from Stephane de Barros at an earlier stage
of this work.
YII and NGG acknowledge support from the National Academy of Sciences of Ukraine by its priority project No.0120U100935 "Fundamental properties of the matter in the relativistic collisions of nuclei and in the early Universe".
JMV acknowledges support from the State Agency for Research of the Spanish MCIU through the ”Center of Excellence Severo Ochoa” award to the IAA (SEV-2017-0709) and project  AYA2016-79724-C4-4-P.
RA acknowledges support from FONDECYT Regular Grant 1202007.

\end{acknowledgements}
%%%%%%%%%%%%%%%%%%%%%%%%%%%%%%%%%%%%%%%%%%%%%%%%%%%%%%%%%%%%%%%%%%%%%%%%%%%%%%%%%
\bibliographystyle{aa}
\bibliography{biblio.bib} 

\end{document}